# Functionalization of graphene sponge electrodes with two-dimensional materials for tailored electrocatalytic activity towards specific contaminants of emerging concern

Elisabeth Cuervo Lumbaque [a,b], Luis Baptista-Pires[a,b], Jelena Radjenovic[a,c]

[a]*Catalan Institute for Water Research (ICRA), Emili Grahit 101, 17003 Girona, Spain*

[b]*University of Girona, Girona, Spain*

[c]*Catalan Institution for Research and Advanced Studies (ICREA), Passeig Lluís Companys 23, 08010 Barcelona, Spain*

** Corresponding author:*

*Jelena Radjenovic, Catalan Institute for Water Research (ICRA), Emili Grahit 101, 17003 Girona, Spain*

Phone: + 34 972 18 33 80; Fax: +34 972 18 32 48; E-mail: jradjenovic@icra.cat

**Highlights**

- 2D materials increased the electrocatalytic activity of graphene sponge electrodes.

- hBN-doped electrode enhanced the π-π interactions of less polar contaminants.

- Borophene-doped electrode enhanced the removal of polar organic contaminants.

- Electrogenerated $^1O_2$ enhanced the oxidation of pollutants with electron-rich moieties.

- Role of surface-bound and dissolved HO• was elucidated in scavenging experiments.

**Abstract**

Low-cost graphene sponge electrodes were functionalized with two-dimensional (2D) materials, i.e., borophene and hexagonal boron nitride (hBN), using a one-step, hydrothermal self-assembly method. Borophene and hBN-modified graphene sponge anode and N-doped graphene sponge cathode were employed for electrochemical degradation of model persistent contaminants of emerging concern in one-pass, flow through mode, and using low-conductivity supporting electrolyte. Functionalization of the reduced graphene oxide (RGO) coating with 2D materials led to specific modifications in the electrode´s electrocatalytic performance and interactions with the target contaminants. For instance, addition of hBN promoted the adsorption of more hydrophobic organic contaminants via van der Waals and π-π interactions. Functionalization of the graphene sponge anode with borophene enhanced the generation of oxidant species such as $H_2O_2$ and $O_3$ and yielded an order of magnitude higher concentration of hydroxyl radicals (HO•) compared with the non-functionalized graphene sponge anode, thus enhancing the removal of target contaminants. Experiments conducted with the selective radical scavengers indicated the key role of surface-bound hydroxyl radicals ($HO•_{ads}$) and singlet oxygen ($^1O_2$) in electrochemical degradation. The study demonstrates that functionalization of graphene sponge electrodes with 2D materials can enhance their electrocatalytic activity and modulate the interaction with specific organic contaminants, thus opening new possibilities for designing electrodes tailored to remove specific groups of pollutants.

## 1. Introduction

The occurrence of contaminants of emerging concern such as pharmaceuticals, personal care products, pesticides, and industrial chemicals in the aquatic environment is a global concern due to their known or suspected ecotoxicological effects and potential impacts on human health. One emerging solution to minimize the presence of contaminants in our water cycle and enhance water security is the introduction of decentralized and distributed (waste)water treatment and reuse systems. However, existing technologies lack practicality and/or struggle to secure cost-effectiveness of water purification and reuse at smaller scale. New technologies for point-of-use and point-of-discharge (waste)water treatment are urgently needed to complement the existing treatment networks, but also for off-grid locations, which is particularly challenging and requires robust, autonomous, and modular systems with small footprint.

Electrochemical processes satisfy many operational and design requirements of a decentralized and distributed (waste)water treatment technology, as they do not entail the use of chemical reagents, do not form a residual waste stream, and are generally robust and versatile systems capable of treating contaminated water of different origin [1,2]. Two major limitations of electrochemical oxidation – formation of toxic chlorinated byproducts in the presence of chloride, and high cost of anode materials – have recently been overcome by the development of new material, graphene sponge electrodes [3,4]. Graphene sponge electrodes are structurally stable and have an estimated cost of less than €50 per $m^2$ of projected surface area, which is several folds lower compared to the commercial boron doped diamond (BDD, ~ €4,200 $m^{-2}$) and dimensionally stable anodes (DSA, ~€3,000 $m^{-2}$) [5]. More importantly, graphene sponge anode displays exceptional electrochemical stability during anodic polarization due to the strong attachment of the reduced graphene oxide (RGO) coating and the employed synthesis template (mineral wool, comprised of silicate fibers) and is the first reported anode material with low electrochemical activity towards chloride. Recent studies demonstrated no chlorate and perchlorate formation even at high current densities, and current efficiency of chlorine formation of less than 0.1% in the presence of high chloride concentration [3,4].

Two-dimensional (2D) materials beyond graphene are in the research spotlight due to their exceptional physical, electrical, chemical, and optical properties. They present ultrahigh specific surface area, abundant exposed active sites, and low charge transfer resistance, and can be combined into so-called van der Waals heterostructures, thus allowing the exploration of new physicochemical and electrochemical effects [6]. Although 2D materials such as hexagonal boron nitride (hBN) and 2D boron, i.e., borophene, have been extensively investigated for enhancing the oxygen evolution reaction (OER) in the energy-related applications [7,8], they have not been applied to date for electro-oxidation of organic pollutants. In this study, we used hBN and homoatomic borophene to tune the structure and electrocatalytic performance of graphene sponge anode, and thus tailor its electrochemical activity with contaminants. Heterostructures of RGO and 2D materials were employed as anodes and coupled with the previously developed N-doped graphene sponge cathode [4] for electrochemical degradation of a set of model contaminants known to be persistent to oxidation by hydroxyl radicals (HO•) and ozone ($O_3$), namely carbamazepine (CBZ), iopromide (IPM), diatrizoate (DTR), triclopyr (TCP) and N,N-Diethyl-m-toluamide (DEET) [9–11]. The objectives of the study were*: i)* to determine the impact of functionalization with 2D materials on the surface and electrochemical properties of the graphene sponge anode, and consequently its reactivity with specific pollutants, *ii)* to investigate the electrocatalytic performance of 2D-modified graphene sponge anodes, *iii)* to determine the main oxidant species contributing to electrochemical degradation of contaminants, and elucidate the role of surface-bound and dissolved oxidants, and *iv)* to investigate the impact of current density on the generation of oxidant species and removal of contaminants. The obtained results demonstrate the feasibility of tailoring the electrochemical activity of graphene sponge anode for the removal of organic pollutants having specific physico-chemical properties.

## 2. Material and methods

### 2.1 Chemicals and reagents

Analytical grade standards for CBZ, IPM, DTR, TCP and DEET, allyl alcohol (AA), *tert*-butanol (*t*-BuOH) and furfuryl alcohol (FFA) were purchased from Sigma-Aldrich. Acetonitrile and HPLC grade water (LC grade) were supplied by Merck. All solvents and chemicals used in this study were of analytical grade.

### 2.2 Fabrication of functionalized graphene sponges

2D materials borophene and hBN were prepared through a liquid-phase exfoliation from bulk boron and boron nitride powders [12,13]. In brief, borophene and hBN were synthesized using 10 g $L^{-1}$ aqueous solutions of boron and boron nitride, respectively, that were sonicated using an ultrasonic homogenizer (Tefic Biotech, 650W) for 4 h. The supernatant was collected for further use. The bulk boron and boron nitride powders were characterized using scanning electron microscopy (SEM) (**Figure S1**) and the presence of 2D nanosheets in the supernatant was verified using laser excitation (**Figure S2**). During laser light scattering, the intensity of the scattered light depends on the size of the particles and fluctuates on very short time scales; smaller particles are displaced further by solvent molecules, move faster and scatter light at larger angles [14]. The solutions employed for the synthesis of borophene-RGO and hBN-RGO sponges were obtained by adding 15 mL of the respective supernatants to 65 mL of GO dispersion (4 g $L^{-1}$) and stirring for 6 h at room temperature. Then, mineral wool template was soaked in the obtained solution and subject to a previously developed one step hydrothermal synthesis method (12 h at 180ºC) [3,4]. The supernatants and resulting graphene-based sponges were characterized using SEM, X-ray photoelectron spectroscopy (XPS), X-ray diffractometry (XRD), and zeta (ζ) potential measurements. Details of the employed surface characterization techniques are summarized in **Text S1**. The synthesis and characterization of N-doped graphene sponge employed as cathode is reported in our previous study [4]. Briefly, to introduce N-dopants into the RGO coating, a solution consisting of 42 g of urea dissolved in 4 g $L^{-1}$ GO solution was used to soak the mineral wool, followed by the hydrothermal synthesis for 12 h at 180ºC. Functionalized graphene sponges were connected to stainless steel current feeders and subject to electrochemical characterization via electrochemical impedance spectroscopy (EIS) and cyclic voltammetry (CV) analysis using a multi-channel potentiostat/galvanostat VMP-300 (BioLogic). The EIS experimental data was fitted using the BioLogic EC-lab software, as describe in our previous study [4]. The electrochemical active surface area ($S_E$) was calculated according to Ali et al [15], where the imaginary parts in the impedance data were determined using phosphate buffer (1 M, pH 7) for the frequency range 50 kHz-0.01 Hz at open circuit potential of 4 mV.

### 2.3 Electrochemical experiments

All experiments were conducted in a cylindrical reactor made of methacrylate-based polymers [4], in one pass flow-through mode at a flow rate of 5 mL $min^{-1}$, resulting in a hydraulic residence time (HRT) of 3.45 min. Borophene or hBN-functionalized RGO anode and N-RGO cathode were separated by a fine polypropylene membrane and pressed against each other to minimize the reactor dead volume, and the applied flow direction was hBN- or borophene-RGO anode/N-RGO cathode. The projected surface area of each electrode was 17.34 $cm^2$, thus the normalized volumetric flux was 175 L $m^{-2}$ $h^{-1}$ (LMH). Leak-free Ag/AgCl reference electrode (Harvard Apparatus) was sandwiched between the electrodes and isolated with the polypropylene membrane to avoid short-circuiting. A stock solution containing all target contaminants (CBZ, IPM, DTR, TCP and DEET) was added to the supporting electrolyte (10 mM phosphate buffer, pH 7.2, 1.1 mS $cm^{-1}$) at the initial concentration of 1 µM of each contaminant. Samples were collected at the reactor exit at predetermined bed volumes, and 900 µL solution was quenched immediately with 100 µL of methanol. Before applying the current, open circuit (OC) experiments were conducted to verify the loss of target contaminants due to their adsorption onto the graphene-based sponge electrodes. Next, experiments were performed in chronopotentiometric mode using a VMP-300 potentiostat (Biologic) with the stepwise increase in the applied anodic current, i.e., 75, 200 and 400 mA, resulting in the anodic current densities of 43, 115 and 231 A $m^{-2}$, respectively, calculated using the projected anode surface area. Terephthalic acid (TA) was used to evaluate the formation of HO• radicals at 43, 115 and 231 A $m^{-2}$ (20 mg $L^{-1}$ TA, 10 mM phosphate buffer, pH 7.2). TA reacts rapidly HO• ($k_{TA,HO\cdot} = 4\times10^9$ $M^{-1}$ $s^{-1}$), does not react via direct electrolysis [16], has very low reactivity with ozone [17], does not adsorb onto the graphene sponges and has very low reactivity with singlet oxygen ($^1O_2$), with the bimolecular rate constant, $k_{TA,1O2} << 10^4$ $M^{-1}$ $s^{-1}$ [18].

The quasi steady state HO• concentration, $[HO\bullet]_{SS}$ was determined as a ratio of pseudo-first rate constant of TA decay ($k_{TA}$, $s^{-1}$) and $k_{TA,HO\cdot}$ [19].

The role of surface-bound and dissolved HO• in the electrochemical degradation of model contaminants was further probed using selective radical scavengers, AA and *t*-BuOH, added at concentrations of 100 mM (excess concentration of scavengers) and 5 µM (molar concentration equivalent to the total initial concentration of target contaminants) in the experiments conducted at the highest applied current density (231 A $m^{-2}$). Likewise, the role of electrochemically generated $^1O_2$ was probed using FFA in excess (20 mM) and at low 5 µM concentration. FFA reacts with $^1O_2$ with a reaction rate constant, $k_{FFA,1O2} = 1.2\times10^8$ $M^{-1}$ $s^{-1}$ [20]. High concentration of scavengers was added to exclude the contribution of both surface-bound and dissolved oxidants, whereas trace concentration of scavengers was added to investigate the role of surface-bound HO• and $^1O_2$ in conditions of competitive kinetics, and without saturating the electrode surface with alcohol that may block the electrode´s active sites. All experiments were performed in triplicate, and the results are presented as mean values with their standard deviations. The recorded potentials were corrected for ohmic drops, calculated based on the EIS data and fitted using the BioLogic EC-lab software.

## 2.4 Analytical methods

Target contaminants were analyzed in multiple reaction monitoring (MRM) mode by 5500 QTRAP hybrid triple quadrupole linear ion trap mass spectrometer (QLIT-MS) with a turbo Ion Spray source (Applied Biosystems), coupled to a Waters Acquity Ultra-Performance™ liquid chromatograph (UPLC). Details of the analytical methods were summarized in **Text S2** and **Table**

**S1 and S2**. The concentration of the electrogenerated $H_2O_2$ was determined by a spectrophotometric method [21] with the quantification limit of 1 μM. The concentration of ozone was measured by the standard indigo method with the quantification limit of 0.4 μM [22]. The concentration of TA used to estimate the steady-state concentration of HO· was determined using HPLC-UV (Agilent Technologies 1200 series) analysis at 239 nm (quantification limit of 1.98 mg $L^{-1}$). The presence of GO and RGO in the treated water was verified by measuring the absorbance at wavelengths of 231 nm and 253-270 nm using UV-Vis diode array spectrophotometer (Agilent Technologies), according to the previously published methods [23,24].

## 3. Results

### 3.1 Surface and electrochemical characterization of 2D materials and 2D-functionalized graphene sponges

The size and morphology of the initial boron and boron nitride were characterized by SEM (**Figure S1**), whereas the resulting 2D materials synthesized by the liquid exfoliation, and graphene sponges with the incorporated 2D materials are presented in **Figures 1A-D**. The initial boron and boron nitride present a granular and disk-like shape morphology in large micrometric clusters, respectively (**Figure S1**). Upon the employed exfoliation process, nanometric-sized 2D materials borophene (**Figure 1A**) and hBN (**Figure 1C**) are formed, as corroborated by the Tyndall effect on the exfoliated solutions (**Figure S2**). After the hydrothermal process using a mixture of the exfoliated solutions, GO and mineral wool, high-resolution SEM image revealed successful incorporation of 2D materials in the final inner and outer RGO structure (**Figure 1B**, **D**), creating exposed active sites at the graphene sponge electrode surface.

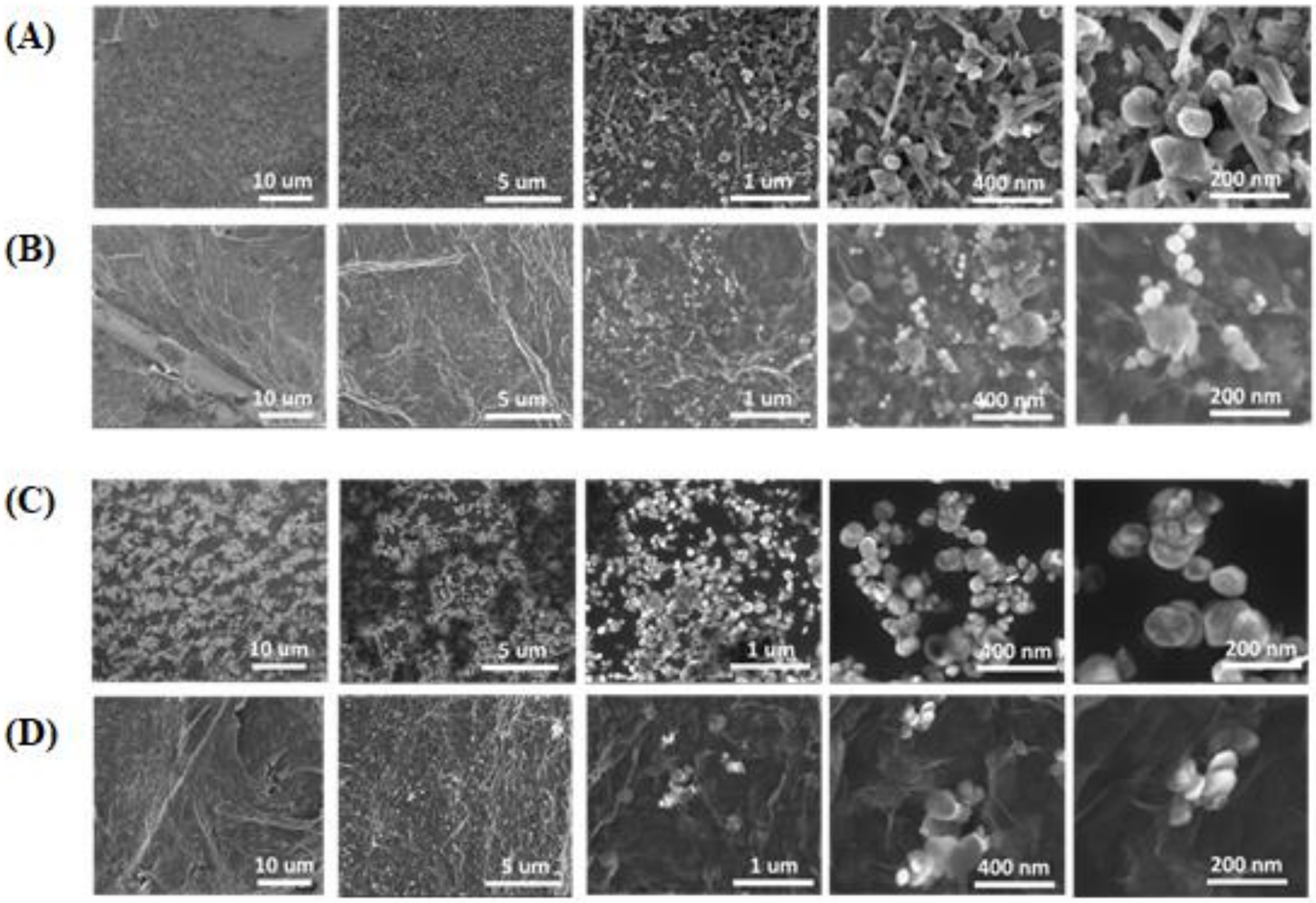

**Figure 1.** Scanning electron microscopy (SEM) images of: **A)** borophene, **B)** borophene-RGO, **C)** hBN, and **D)** hBN-RGO.

Effective hydrothermal reduction of graphene was confirmed by the XPS analysis, with a C/O ratio increased from 1.7 of the initial GO solution to C/O ratio of 2.5 for the RGO coating (**Table S3**). XPS analysis was also used to determine the effect of the exfoliation process on the resulting 2D materials (**Figure 2 and Figure S3**). Borophene has a higher atomic oxygen content compared with the starting solution of bulk boron, because of the sonication process; the peak located at 533 eV is the dominant functional bounding of borophene and it is attributed to the C=O, C-N, and B-O bonds (**Figure 2A**, **Table S4**). Introduction of oxygen functional groups is confirmed by the higher % of B-O bonds (11.4%; 188 eV) and lower % of B-B bonds (30.7%; 187.6) compared with boron (5% B-O and 45.9% B-B bonds) (**Table S4**). Similar results of XPS analyses were obtained for the synthesized hBN, with the higher % of B-O bonds (39.8% B-O and 1.5% $B_2O_3$) and lower % of B-B and B-N bonds (1.8% in both cases) compared with the boron nitride (6.1% B-O, 1.4% $B_2O_3$, 3.2% B-B bonds and 29.2% B-N). Also, signal intensity of the peaks of oxygenated carbons such as C=O, C-N, and B-O was increased after exfoliation (**Figure 2B**). XPS analysis cannot detect the presence of 2D materials in the graphene sponge electrodes because of the low concentration of borophene and hBN in the sample, and limited sensitivity of the XPS analyses due to the mineral wool interference (example in **Figure S4**). Nevertheless, the presence and impact of 2D materials on the graphene sponge anode performance was clearly observed in their electrochemical characterization, as well as experiments conducted with the target pollutants, as explained further in the text.

**(A)**

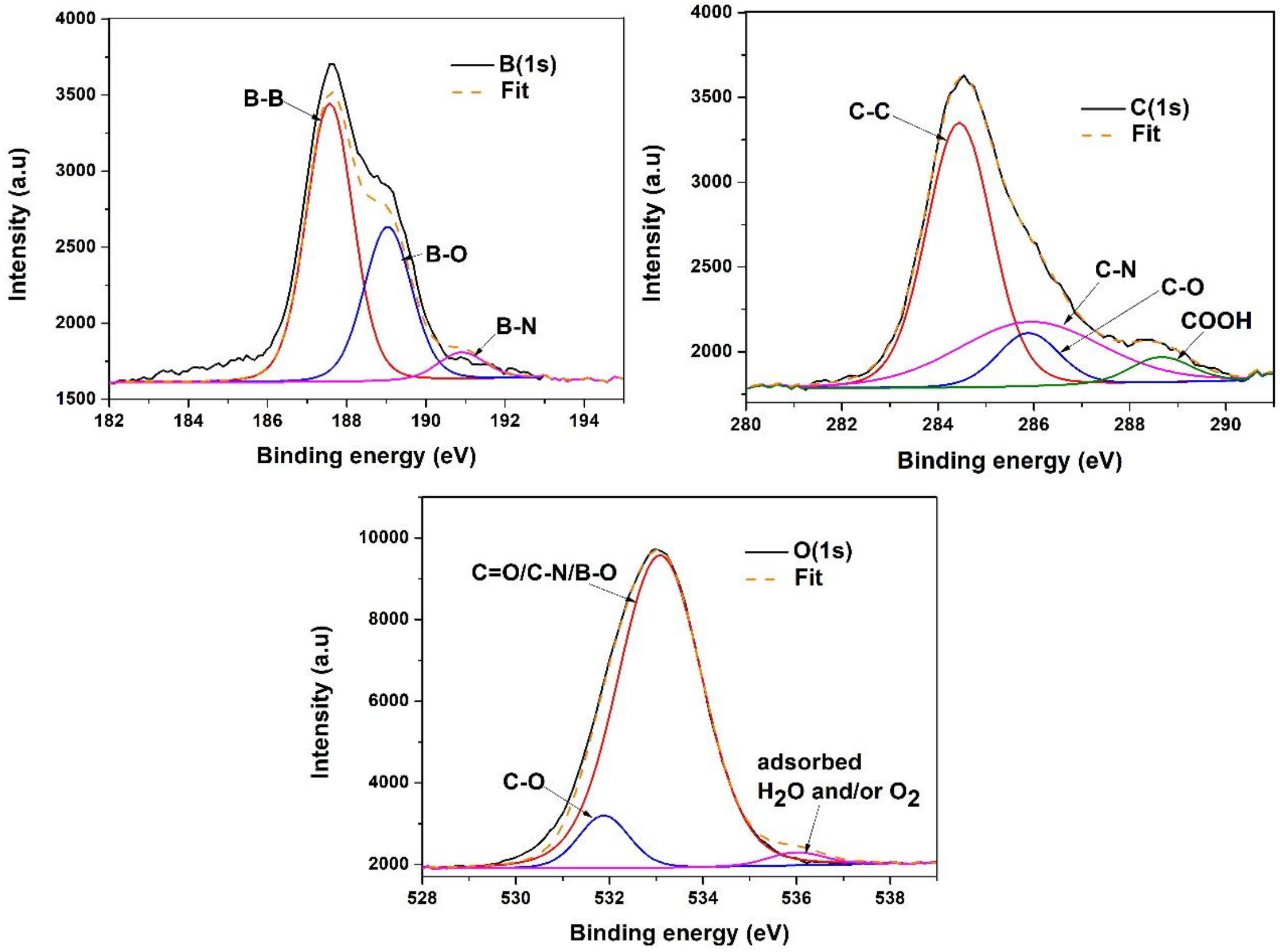

(B)

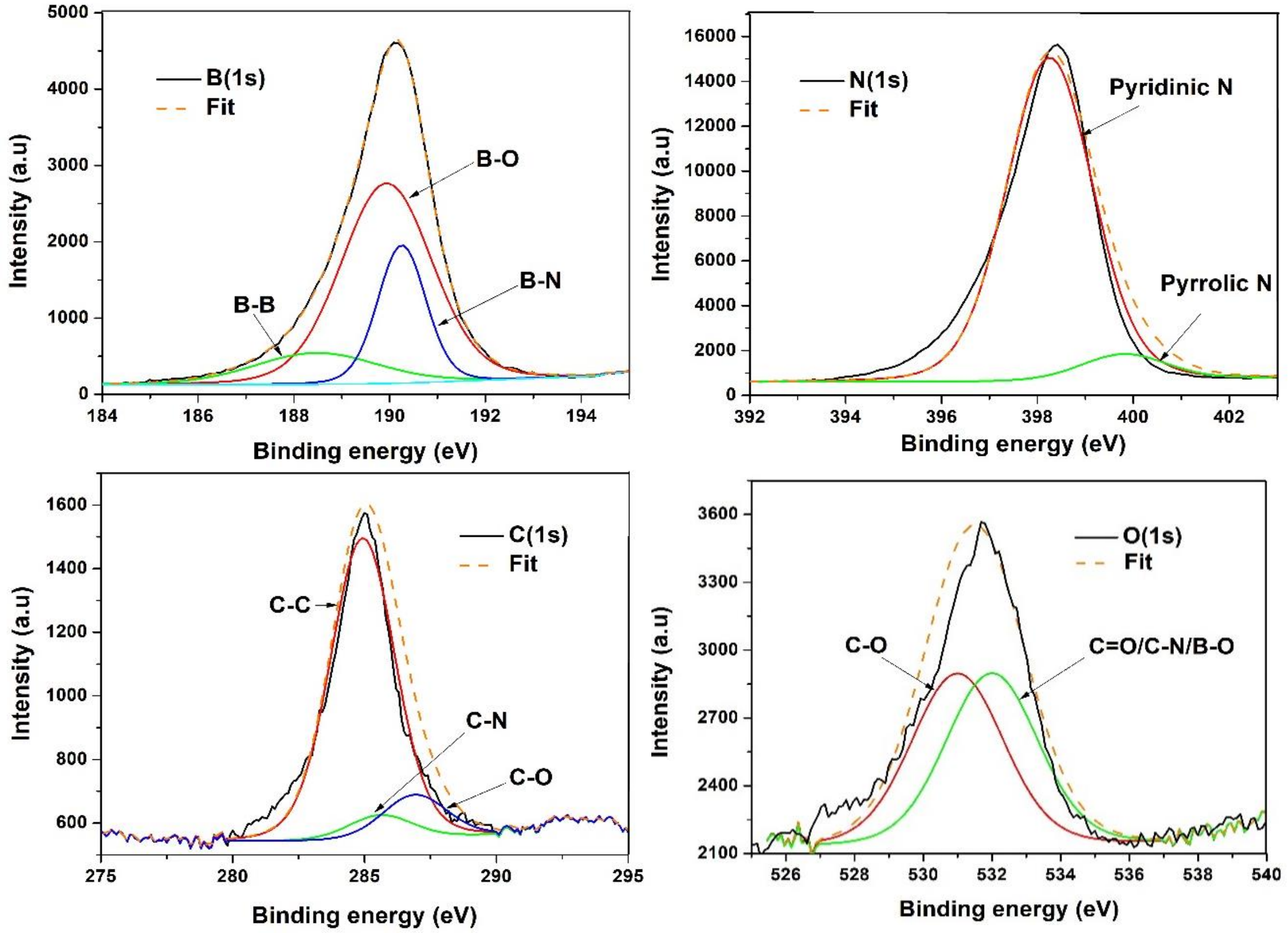

**Figure 2.** X-ray photoelectron spectroscopy (XPS) analyses of supernatants: **A)** borophene and **B)** hBN.

XRD patterns indicated the amorphous nature of the borophene supernatant, which may be due to the geometrical frustration caused by the positions of terminal B-H bonds located on the B-H-B bridging bonds [25]. In contrast, for hBN supernatant, XRD pattern indicated that most of the crystals are oriented in the (002) direction, which is a characteristic peak observed at 26.7° (**Figure S5**). The surface charge of the graphene sponge electrodes and electron density of the active sites will impact their electrostatic interaction with organic contaminants both in the absence of current (i.e., OC) and during current application [3,26]. The determined $\zeta$ potentials of the RGO, borophene-RGO and hBN-RGO coating were -18.1 mV, -33.9 mV and -34.9 mV, respectively. The negative charge of borophene and hBN nanosheets was previously reported in literature [27,28] and leads to more negative surface charges of heterostructures of RGO and 2D materials compared with the unmodified RGO.

EIS and CV measurements were carried out to evaluate the charge transfer resistance of different graphene sponge anodes, and their electrochemical activity towards chloride, respectively. The near-vertical slope in the EIS analysis at low frequency was used for obtaining the imaginary part of the double-layer capacitance (Cdl), where Cdl represents the real capacitance of the cell at the corresponding frequency [29]. The calculated values of Cdl were 0.067 F $g^{-1}$ (RGO), 0.081 F $g^{-1}$ (hBN-RGO) and 0.90 F $g^{-1}$ (borophene-RGO) (**Figure S6**). Electrochemically active surface area ($S_E$) was calculated as $S_E$=Cdl/Cd [15], where Cd is a constant value of carbon-based electrode (10 μF $cm^{-2}$) [30]. The calculated values of $S_E$ were 0.67 $m^2$ $g^{-1}$ for RGO, 0.81 $m^2$ $g^{-1}$ for hBN-RGO, and 0.90 $m^2$ $g^{-1}$ for borophene-RGO. These values are somewhat lower compared with the previously determined BET specific surface area of 1.4 $m^2$ $g^{-1}$ for graphene sponges [4] and indicate that only a portion of the sponge electrode surface is electrochemically active, which may be caused by its high hydrophobicity (contact angle of >137º) and uneven wetting of the sponge surface. **Figure 3A** illustrates the Nyquist plots, with resistances normalized with the electrochemically active surface area of the unmodified graphene sponge electrode (i.e., RGO electrode) and 2D-functionalized graphene sponges (hBN-RGO and borophene-RGO electrode). Borophene-RGO exhibits a shorter diffusion length (near-vertical slope) than hBN-RGO and RGO anodes, indicating a faster response of the system attributed to enhanced diffusion of ions within the material. In addition, borophene-RGO and hBN-RGO showed more vertical slope than the RGO anode, further confirming their higher capacitance. The semi-circle in Nyquist plots represents a charge transfer resistance ($R_{ct}$) at the electrode-electrolyte interface. Smaller diameters of the semi-circle for borophene-RGO and hBN-RGO electrodes indicates faster electron transfer compared with the undoped graphene sponge anode. The electrical resistance ($R_{ct}$) was decreased from 842,535 $\Omega \cdot cm^2$ to 583,254 $\Omega \cdot cm^2$ and 487,018 $\Omega \cdot cm^{-2}$ by incorporating hBN and borophene, respectively, into the RGO coating. Doping of RGO with heteroatoms such as boron, nitrogen and others increases the density of free charge carriers and enhances the electrical conductivity of the material [31]. Borophene is electron-deficient [32] and its addition to the RGO coating creates defects in the RGO structure, thus improving the charge carrier density and the electrocatalytic activity of the material. In the case of hBN, this result may be surprising as hBN is an insulator. However, previous studies reported an enhancement in the electrocatalytic activities of hBN-functionalized RGO for the optimum amount of added hBN [33,34]. This was explained by the strong graphitic interactions of hBN with RGO and formation of new active sites (-O-C-B-N-) with effective charge distribution on C and B atoms. Furthermore, the presence of hBN nanosheets makes it difficult to stack many layers of RGO nanosheets completely, allowing a maximum exposure of active sites in hBN-RGO electrode [33].

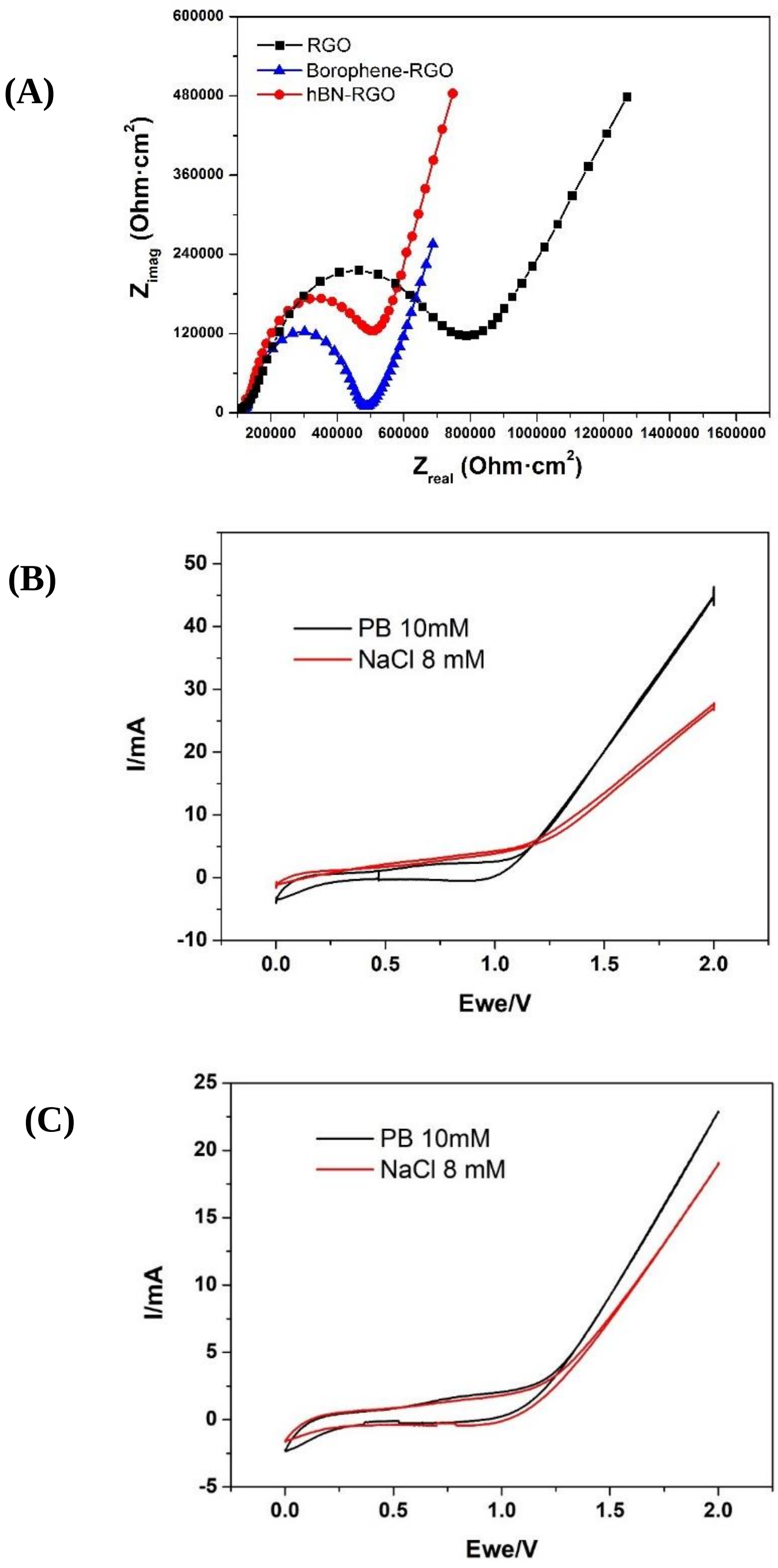

**Figure 3. A)** Nyquist plots of RGO, borophene-RGO and hBN-RGO anodes in 10 mM phosphate buffer (PB), and CV scans recorded in 10 mM PB and in 8 mM NaCl solution, both at pH 7, 1.2 mS $cm^{-1}$, and using 10 mV $s^{-1}$ scan rate for: **B)** borophene-RGO anode, and **C)** hBN-RGO anode, all coupled to N-RGO cathode.

From the obtained CVs in the 10 mM phosphate buffer (pH 7, 1.1 mS $cm^{-1}$) and 8 mM NaCl solution (pH 7, 1.1. mS $cm^{-1}$), it is evident that both 2D-functionalized graphene sponge anodes have no reactivity towards chloride oxidation (**Figure 3B, C**), in accordance with our previous studies [3,4]. This result stands in stark contrast to the high electrocatalytic activity of commercial electrodes such as BDD and DSAs towards chlorine formation, for which the presence of $Cl^-$ leads to a pronounced increase in current and shift of the onset potential to less positive values in the anodic sweep of the CV, due to the chlorine evolution reaction ($2Cl^- \rightarrow Cl_2 + 2e^-$, $E^o$=1.36 V/SHE) [35,36]. Given this pronounced electrochemical intentness towards $Cl^-$, graphene sponge anodes also do not produce any chlorate and perchlorate even at high current densities and in the presence of high chloride concentrations, representing a major advantage over other anode materials [4].

### 3.2 Electrooxidation of persistent organic contaminants

**Figure 4A-C** shows normalized effluent concentrations in the initial ($OC_i$) and final OC ($OC_{final}$), and at different applied current densities using RGO, borophene-RGO and hBN-RGO anode, all coupled to N-RGO cathode. Functionalization of graphene sponge anode with hBN enhanced the adsorption of CBZ and DEET, and their removal efficiencies in the OC were doubled, i.e., from 22±1% to 47%±1.2% (for CBZ) and from 11±1% to 22%±2% (for DEET) using RGO and hBN-RGO anode, respectively (**Figure 4B**). CBZ and DEET have the highest logD values among the investigated contaminants (logD=2.3 and 2.2, respectively, **Table S5**), and are thus more likely to adsorb at the graphene sponge surface via van der Waals interactions. On the other hand, the presence of hBN did not impact the removal of more polar IPM, DTR and TCP in the OC (<10% removal). hBN has been extensively investigated for the adsorption of organic pollutants, solvents and oils [37] and its high adsorption capacity is explained by the π–π stacking between the aromatic

ring moieties of organic pollutants and the hexagonal cells of hBN [38]. On the other hand, functionalization of graphene sponge anode with borophene did not impact the adsorption of CBZ, DEET and TCP, yet it led to an enhanced removal of highly polar IPM and DTR in the OC, from 12±2% and 8±1% at RGO anode to 21±1% and 14±1% for borophene-RGO anode, respectively. Both hBN-RGO and borophene-RGO had very similar determined ζ potentials (-33.9 mV and -34.9 mV, respectively). However, whereas hBN promotes π-π interactions, borophene active sites are dominated by the electrostatic interactions. IPM dipole (IPM is uncharged at pH 7) and negatively charged DTR have significantly higher polar surface areas compared with other target contaminants (**Table S5**), resulting in more pronounced electrostatic interactions due to higher molecular electrostatic potentials [39].

The removal of persistent organic contaminants due to both adsorption and electrooxidation showed very good repeatability with low deviations **(Figure 4A-C).** The stability of the graphene sponge was assessed by applying up to 231 A $m^{-2}$ of the anodic current density and verified by the analysis for the presence of GO and RGO in the reactor effluent, which showed that both GO and RGO were below limit of detection (LOD=0.0176 mg $L^{-1}$) (**Figure S7**). In addition, stable electrode potentials were recorded in all experiments (**Table S6 and Figure S8**). Ohmic-drop corrected anode potentials were in the range 2.5-5.3 V/SHE for the applied current densities (**Table S7**). The energy consumed with a current density of 231 A $m^2$ and effluent flux of 175 LMH was 5.73 kWh $m^{-3}$ for hBN-RGO/N-RGO and 4.13 kWh $m^{-3}$ for borophene-RGO/N-RGO. In both hBN-RGO/N-RGO and borophene-RGO/N-RGO systems, stepwise increase in anodic current density led to a proportional increase in the observed removal efficiencies of all target contaminants (**Figures 4B, C**). Furthermore, functionalization of graphene sponge anode with both borophene and hBN yielded higher removal efficiencies for all target contaminants (**Figure 4A**). For example, removal efficiency of DTR at 231 A $m^{-2}$ was increased from 60±0.3% using undoped graphene sponge anode (RGO/N-RGO system), to 76±1% and 89±1% using hBN- and borophene-doped graphene sponge anode, respectively. IPM exhibited similar behavior, with 57±0.2%, 78±2% and 95±0.1% removal at 231 A $m^{-2}$ using RGO, hBN-RGO and borophene-RGO anode, respectively. Due to the high polarity of IPM and DTR and thus their limited removal in the OC, improved removal with the application of current at 2D-functionalized graphene sponge anodes was attributed to the enhanced electrocatalytic activity of anode, yielding higher amounts of electrogenerated oxidants and thus their improved degradation, as explained further in the text.

DEET and CBZ, having the highest logD values among the target pollutants, were removed to a larger extent at the hBN-doped graphene sponge anode (42±2% and 79±2% at 231 A $m^{-2}$) compared with RGO (22±0.1% and 60±0.3%) and borophene- RGO anode (31±0.9% and 71±2 %, respectively). Enhanced adsorption of these contaminants to the hBN-modified surface contributed to their higher electrochemical removal efficiency during the application of current. In the case of TCP, hBN- and borophene-RGO showed very similar performance, with the removal efficiencies in the range 64-64%, 72-68% and 73-71% at 43, 115 and 231 A $m^{-2}$, respectively. These values compare favorably to the TCP removal using non-functionalized RGO anode (31±1%, 40±2% and 46±1% at 43, 115 and 231 A $m^{-2}$, respectively, **Figure 4A**). TCP is a polar compound (logD= -1, **Table S5**), but has high affinity towards π-π interactions due the presence of chloride substituents [41], and the presence of N heteroatom in pyridine, which reduces the spatial extent of the π-electron cloud and favors the π-π stacking of TCP [42]. Thus, the presence of hBN likely promoted the interaction of TCP with the graphene sponge surface, whereas the presence of borophene leads to an enhanced generation of oxidants and thus its improved removal.

The lowest removal efficiencies were obtained for highly persistent DEET in all three systems, with the hBN-RGO/N-RGO system achieving the highest DEET removal at 231 A $m^{-2}$, i.e., 42±2%, due to the improved π-π interactions of DEET with the hBN nanosheets in the RGO coating. Although the investigated systems did not achieve complete removal of the persistent target pollutants at the highest applied current, these results were obtained using low conductivity electrolyte (1.1 mS $cm^{-1}$), employed to evaluate the system performance under the conditions of high ohmic drop of real contaminated water. A feasible strategy to improve the electrochemical removal of persistent contaminants is to stack up several graphene sponge electrodes, or use a larger electrode, e.g., with 10x10 cm projected surface area typically used in electrochemical modules, which should perform better compared with the smaller graphene sponge electrodes employed in this study (4.5 cm diameter, 0.5 cm height). Given that the estimated cost of graphene sponges is less than 50€ per $m^2$ of the projected surface area [4] and their inherent compressibility and flexibility, multiple reactor geometries with high electrode surface area to reactor volume ratio are possible.

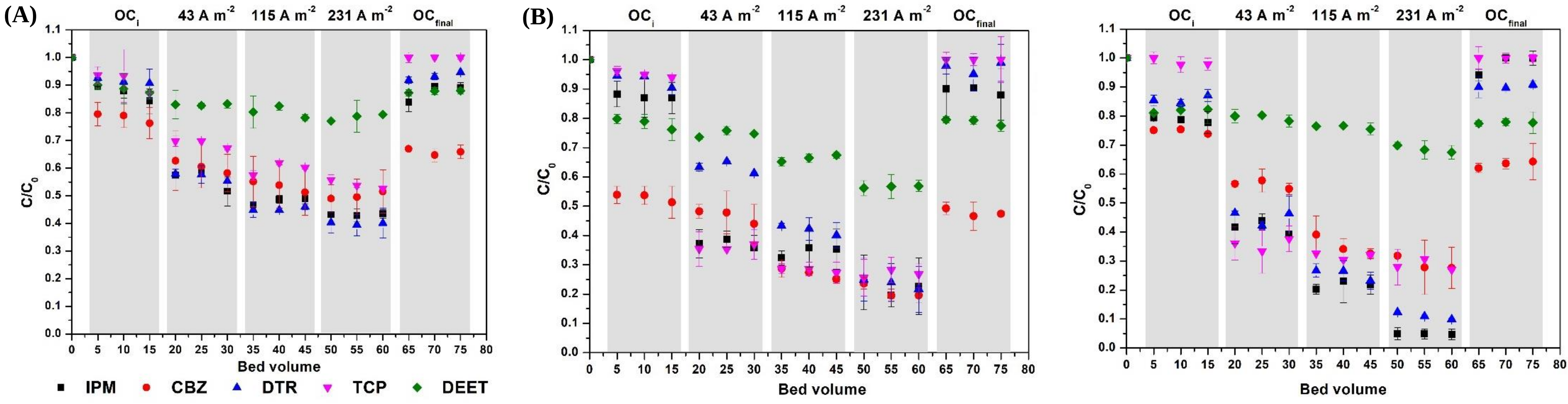

**Figure 4.** Electrochemical removal of model contaminants using: **A)** RGO anode, **B)** hBN-RGO anode, and **C)** borophene-RGO anode, all coupled to N-RGO cathode, in the initial open circuit ($OC_i$), in chronopotentiometric runs at increasing current densities, and in the final OC ($OC_{final}$).

### 3.3 Analysis of the electrogenerated oxidant species

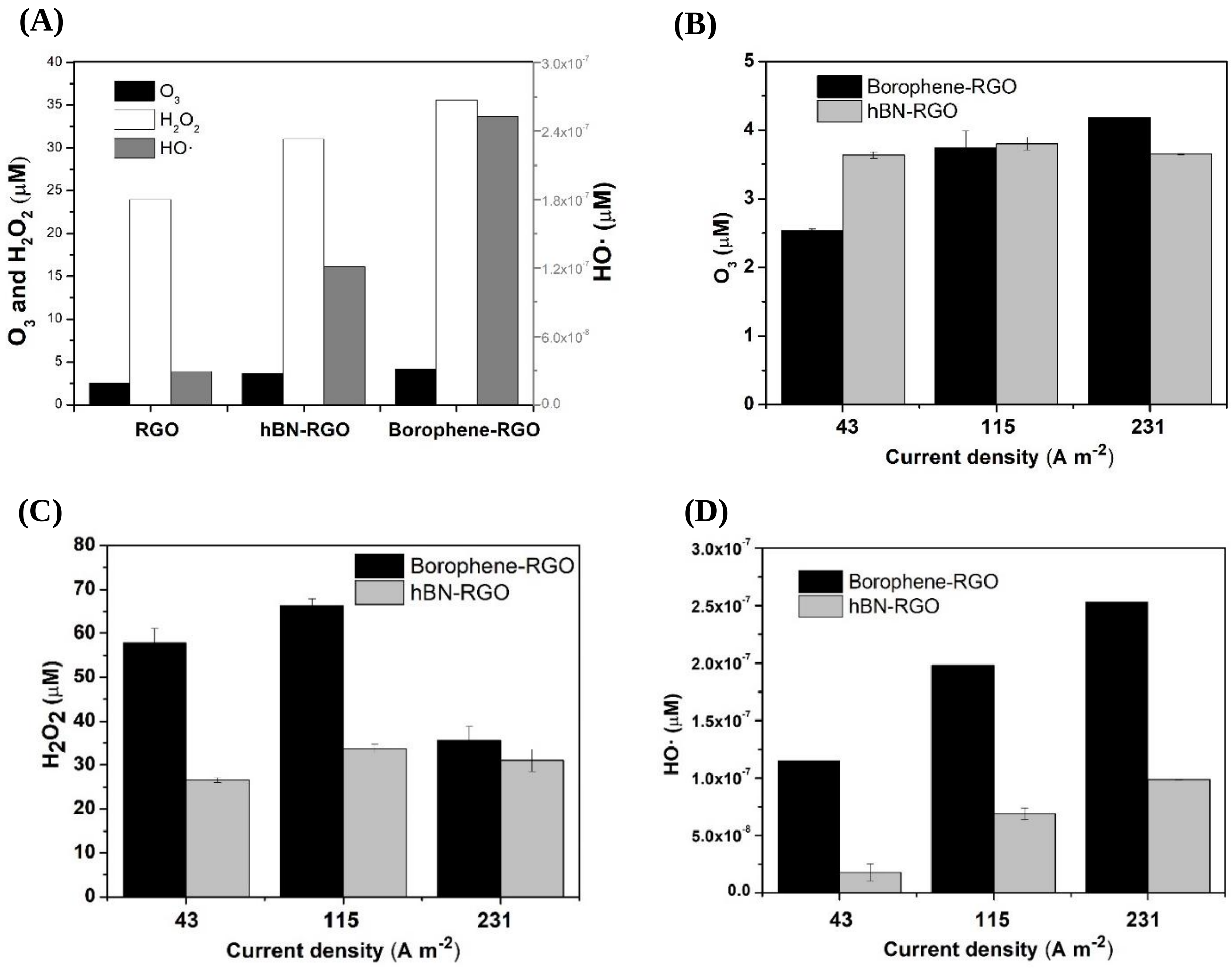

**Figure 5.** Concentrations of oxidants measured in flowthrough reactor using different graphene sponge anodes, all coupled to N-RGO cathode: **A)** ozone ($O_3$), hydrogen peroxide ($H_2O_2$) and hydroxyl radical ($^{\bullet}OH$) measured for RGO, borophene-RGO and hBN-RGO anode in steady-state operation at 231 A $m^{-2}$, and **B)** $O_3$, **C)** $H_2O_2$ and **D)** $^{\bullet}OH$ radicals produced in steady-state chronopotentiometric experiments at 43, 115 and 231 A $m^{-2}$ using borophene-RGO and hBN-RGO anode. To determine the amount of anodically generated $O_3$, its concentration was measured in cathode-anode flow direction. The concentrations of $^{\bullet}OH$ and the cathodically generated $H_2O_2$ were determined in the anode-cathode flow direction employed in all electrochemical degradation experiments.

Electrochemical degradation of pollutants can proceed via direct electron transfer to the anode and by the electrochemically generated oxidants such as $O_3$ and HO•. Ozone was formed in low quantities when using undoped graphene sponge anode (**Figure 5A**) (0.12 mg $L^{-1}$ at 231 A $m^{-2}$), with a moderate increase to 0.18-0.20 mg $L^{-1}$ $O_3$ for hBN- and borophene-RGO anode, respectively (**Figure 5B**). Previous study reported synergetic action of B- and N-active sites in B,N-co-doped mesoporous carbon in the anodic generation of ozone [43]. Similar mechanisms could be at play in the case of borophene-RGO anode, as according to the XPS analysis (**Table S3**) N1s was identified in the sample (1.7 %), originating from the commercial GO solution employed. Nevertheless, given their recalcitrance to ozone (**Table S8**), oxidation of the target contaminants by the anodically generated $O_3$ is unlikely.

The measured amount of $H_2O_2$ was 0.8, 1.1 and 1.2 mg $L^{-1}$ (i.e., 24±1.8, 31.1±2.8, and 35.6 ±3.2 μM) when using RGO, hBN-RGO and borophene-RGO anode at 231 A $m^{-2}$ (**Figure 5A**). 2D materials such as hBN and borophene have been extensively investigated for accelerating the OER [44]. Enhanced anodic generation of $O_2$ would thus yield higher amounts of $H_2O_2$ generated at the N-RGO cathode. Significantly higher amounts of $H_2O_2$ were determined at lower current densities for borophene-RGO anode, i.e., 1.97±0.1 and 2.25±0.1 mg $L^{-1}$ at 43 and 115 A $m^{-2}$ (**Figure 5C**), approximately two times higher compared with the hBN-RGO anode (0.90 and 1.15 mg $L^{-1}$, respectively). This may be due to a larger accumulation of $O_2$ bubbles in the borophene-RGO/N-RGO system at higher currents, which may have decreased the electrochemical activity of the cathode. It should be noted here that N-active sites at the N-RGO cathode activate $H_2O_2$ to HO• [45], and the determined $H_2O_2$ thus represents residual amount of oxidant in the system.

Experiments with TA revealed a drastically enhanced generation of HO· at the 2D-functionalized graphene sponge anodes (**Figure 5D**). Steady-state concentrations of HO· at 231 A $m^{-2}$ were increased for an order of magnitude, from 2.94 x $10^{-14}$ M at RGO anode, to 1.21 x $10^{-13}$ M and 2.53 x $10^{-13}$ M at hBN-RGO and borophene-RGO anode, respectively. The estimated amounts of the electrogenerated HO· are considered conservative, since the exact number of HO· reacting with TA is unknown. When normalized to the applied volumetric flux of 175 LMH, steady-state concentrations of HO·, $[HO\cdot]_{SS}$, at 231 A $m^{-2}$ (ohmic drop-corrected anode potential, $E_{AN}$=3.1 V/SHE) correspond to 1.75x$10^{-10}$ mol $m^{-2}$, 3.88x$10^{-10}$ mol $m^{-2}$ and 1.5 x$10^{-9}$ mol $m^{-2}$ at RGO, hBN-RGO and borophene-RGO anode, respectively. In our previous study, the estimated $[HO\cdot]_{SS}$ for boron-doped graphene sponge anode was 5.77x$10^{-10}$ mol $m^{-2}$ using the same experimental set-up with N-RGO cathode at 173 A $m^{-2}$ (ohmic drop-corrected $E_{AN}$= 3.7 V/SHE) [4]. The change from atomic boron to 2D form of boron, borophene, led to an order of magnitude increase in the amount of electrogenerated HO·, thus evidencing the potential of 2D materials for modulating the electrooxidation activity of graphene-based materials.

The reported bimolecular rate constants of the target pollutants with HO· are in the order of $10^8$-$10^9$ $M^{-1}$ $s^{-1}$ (**Table S8**). Thus, improved the degradation of the target contaminants when using 2D-functionalized graphene sponge anodes can be explained by the enhanced formation of HO· radicals. Electrogeneration of HO· in the investigated flow-through electrochemical system can proceed via surface-based activation of $H_2O_2$ at N-active sites of the N-RGO cathode, decomposition of the anodically generated $O_3$ to HO·, and in the reaction of anodically generated $O_3$ and cathodically produced $H_2O_2$. The amount of $H_2O_2$ measured for three systems was very similar (0.8-1.2 mg $L^{-1}$), suggesting that the difference in HO· generation stems from the improved anodic catalysis with the addition of 2D materials. Direct electrolysis of water at the highly hydrophobic graphene sponge is unlikely to form physiosorbed HO• that can participate in contaminant oxidation. Yet, this cannot be ruled out as electrochemical properties of the anodically polarized graphene-based materials are still poorly understood. The interpretations of water electrolysis mechanisms at conventional anode materials (e.g., BDD, DSA) may fall short for graphene-based anodes, as evident from their surprisingly low electrocatalytic activity towards $Cl^-$ oxidation.

## 3.4 Distinguishing between surface-bound and dissolved oxidants for borophene-RGO/N-RGO system

To distinguish between the participation of surface-bound and dissolved HO•, experiments were performed with the borophene-RGO/N-RGO system in the presence of AA and *t*-BuOH. AA was employed in literature as a probe for surface-bound oxidants because of the strong interaction of its π-orbitals with the electrode surface [46,47], whereas saturated alcohols such as *t*-BuOH react rapidly with dissolved HO• but do not react readily with surface-bound HO• [48]. Furthermore, experiments were also conducted in the presence of FFA to distinguish the contribution of singlet oxygen ($^1O_2$) [49,50]. **Figure 6** shows the contaminant removal efficiencies obtained in steady-state at 231 A $m^{-2}$ and in the presence of excess oxidant scavengers (i.e., 20-100 mM) and low concentrations of scavengers allowing competitive kinetics with organic pollutants (i.e., 5 µM).

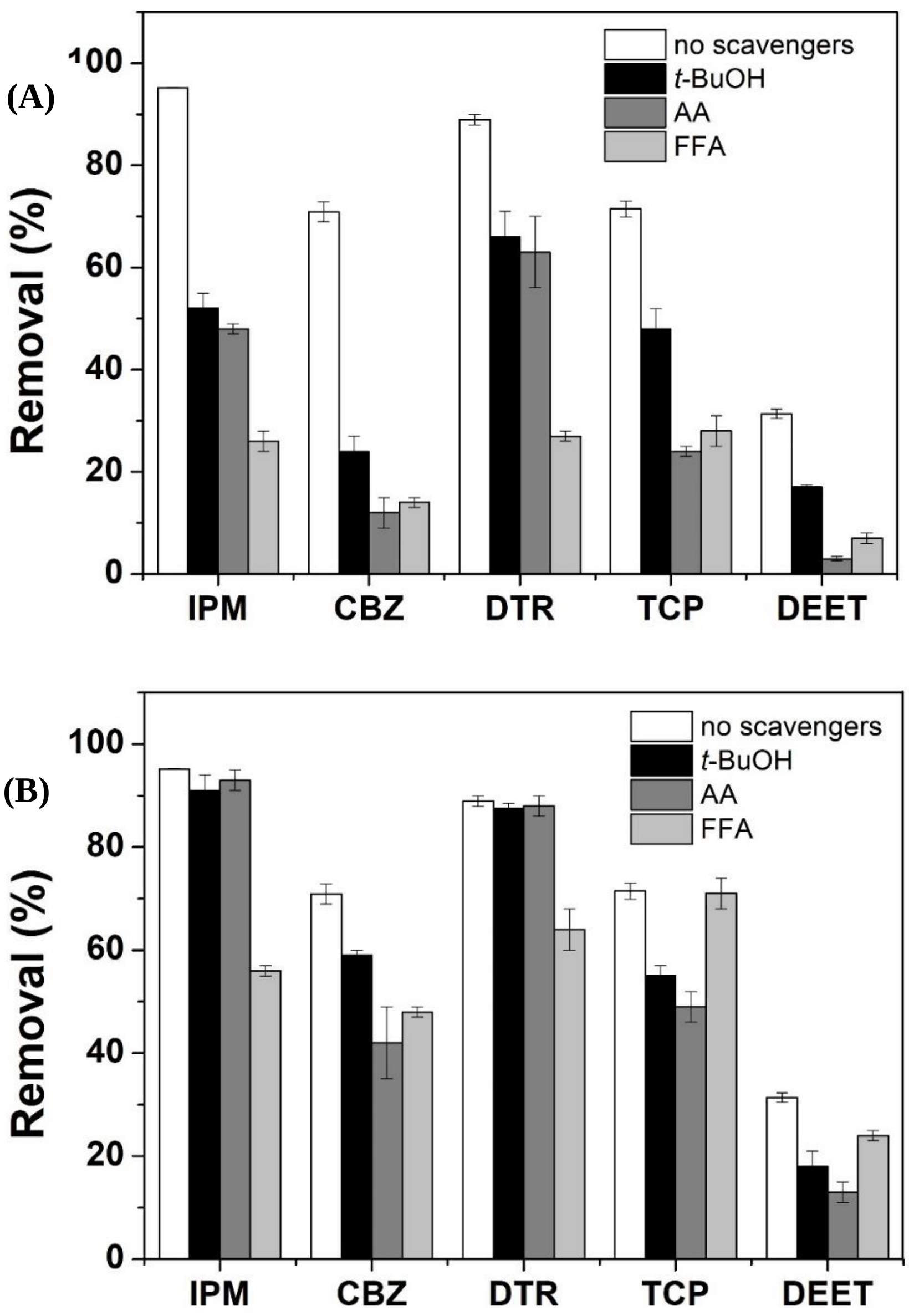

**Figure 6.** Removal of target contaminants, each at the initial concentration of 1 µM, in borophene-RGO/N-RGO system at 231 A $m^{-2}$ in the presence of radical scavengers *tert*-butanol (*t*-BuOH), allyl alcohol (AA) and furfuryl alcohol (FFA) at: **A)** 100 mM concentration, and **B)** 5 µM concentrations.

In the presence of excess AA, *t*-BuOH and FFA, the removal of all target pollutants was significantly inhibited due to the complete scavenging of HO•, both surface-bound HO• ($HO•_{ads}$) and homogeneously formed HO• ($HO•_{bulk}$), as well as $^1O_2$ (**Figure 6A**). Whereas the removal of IPM and DTR was similar in the presence of 100 mM t-BuOH and AA, addition of AA resulted in a two- to five-fold higher inhibition of the removal of DEET, TCP and CBZ. This can be explained by the more pronounced surface interaction of these contaminants compared with highly polar IPM and DTR molecules, and thus more significant role of surface-bound HO•. However, electrodegradation of contaminants was not completely inhibited even for high concentrations of radical scavengers. IPM, DTR, DEET and TCP are recalcitrant to ozone, and CBZ has very low reactivity with $O_3$ ($9.1 \times 10^5$ $M^{-1}$ $s^{-1}$, **Table S8**). This suggests that the oxidation of target contaminants in the presence of excess *t*-BuOH and AA was achieved via direct electrolysis and/or non-radical oxidant, i.e., singlet oxygen ($^1O_2$). When the experiments were run with the addition of excess FFA, a specific scavenger of $^1O_2$, electrodegradation of IPM and DTR was inhibited to a higher degree compared with the runs conducted with excess AA and t-BuOH. For example, the removal of IPM was decreased from 95% in the absence of scavengers, to 52-48% in the presence of excess t-BuOH and AA, and 26% in the presence of excess FFA. These results points out at an important role of $^1O_2$ in the electrodegradation of iodinated contrast media at graphene sponge electrodes. $^1O_2$ has much slower rate constants with most organic pollutants compared with HO• but has a vacant π* orbital and acts as a highly reactive and selective oxidizing species toward electron-rich compounds (e.g., phenols, sulfides, and amines) [51–53]. Thus, $^1O_2$ will react readily with the secondary and tertiary amine groups of IPM and DTR, which are non-protonated at pH 7 (**Table S5**). Compared with the non-selective and short-lived HO•, $^1O_2$ is meta-stable [51,52] and likely to exist not only as surface-bound but also as dissolved species. Most studies focused on electrocatalytic generation of $^1O_2$ are based on the use of $^1O_2$ precursors (e.g., peroxymonosulfate) [54,55]. Direct anodic generation of singlet oxygen has not been demonstrated to date, although $^1O_2$ is suspected to be an intermediate in the oxygen evolution reaction (OER) [56]. In the borophene-RGO/N-RGO system with anode-cathode flow direction, $^1O_2$ can be formed: *i)* in the Haber-Weiss reactions of $HO_2\cdot/O_2\cdot^-$ with $H_2O_2$ (eq. 1, 2), as well as radical–radical recombinations of $HO_2\cdot/O_2\cdot^-$ (eq. 3, 4) [49]; *ii)* in the reaction of $O_2\cdot^-$ with HO· (eq. 5) [57], and *iii)* in the reaction of RGO and anodically generated ozone [58], which was assigned to the unsaturated carbon atoms and defects in RGO acting as active sites for the dissociation of O-O bond [59,60]. It should be noted here that $O_2\cdot^-$, $HO_2\cdot^-$ and HO· are formed by the activation of the cathodically generated $H_2O_2$ at graphite N and pyridinic N-active sites [45].

$$O_2\cdot^- + H_2O_2 \rightarrow {}^1O_2 + HO\cdot + OH^- \qquad \text{(eq. 1)}$$

$$HO_2\cdot + H_2O_2 \rightarrow {}^1O_2 + HO\cdot + H_2O \qquad \text{(eq. 2)}$$

$$2\ HO_2\cdot \rightarrow {}^1O_2 + H_2O_2 \qquad \text{(eq. 3)}$$

$$2\ H^+ + 2\ O_2\cdot^- \rightarrow {}^1O_2 + H_2O_2 \qquad \text{(eq. 4)}$$

$$O_2\cdot^- + HO\cdot \rightarrow {}^1O_2 + OH^- \qquad \text{(eq. 5)}$$

In the case of CBZ, TCP and DEET, FFA had a similar inhibiting effect to the added AA. Although CBZ and DEET can react with $^1O_2$ due to the presence of aromatic rings and double bonds in their structures (**Table S5**), their reactivity is expected to be limited due to the absence of nucleophilic substituents that would have an activating effect on the aromatic ring. In the case of TCP, the presence of three chloride substituents with electron-withdrawing effect will deactivate the aromatic ring, reducing its electron density and resulting in poor reactivity with $^1O_2$. However, excess mM concentrations of oxidant scavengers not only quench the HO· and $^1O_2$, but also inhibit the contaminant interaction with the electrode surface by occupying the electrode´s active sites.

Thus, to further investigate the role of $HO\cdot_{ads}$, experiments were conducted with lower concentration of AA, *t*-BuOH and FFA, i.e., 5µM each, equivalent to the total molar concentration of the target pollutants. This allowed kinetic competition of the scavengers with the target pollutants and better elucidation of the role of oxidant species in their electrochemical degradation. As shown in **Figure 6B**, addition of 5 µM *t*-BuOH and AA lowered the removals of CBZ, TCP and DEET, whereas the removals of IPM and DTR were not affected by the low concentrations of radical scavengers. These results further confirmed the predominant oxidation of the iodinated contrast media by $HO^{\bullet}_{bulk}$, which could not be scavenged to a sufficient degree in the presence of 5 µM of *t*-BuOH and AA. On the other hand, in the conditions of competitive kinetics, electrochemical degradation of CBZ, DEET and TCP was inhibited with the addition of both *t*-BuOH and AA, with the more pronounced effect observed for AA. For example, the removal of CBZ was lowered from 71%, to 59% (with 5 µM *t*-BuOH) and 42% (with 5 µM AA). $HO^{\bullet}$ can be generated at graphene sponge electrodes by the activation of $H_2O_2$ at the N-RGO cathode [61], and decomposition of $O_3$ at the graphene sponge anode [62]. Although AA is expected to react more rapidly with the surface-bound oxidants [46,47], in the case of graphene sponge electrodes, addition of *t*-BuOH also inhibited $HO^{\bullet}_{ads}$. This can be explained by the inhibition of the cathodically generated $HO^{\bullet}_{ads}$ by *t*-BuOH, as tertiary alcohols are pseudo-acidic and more prone to nucleophilic attack compared to allyl alcohol. Furthermore, the reaction of tertiary alcohols with graphene-based materials is favored due to the ability of GO to delocalize and accumulate the negative charge [63]. CBZ and DEET are less polar compounds compared with IPM and DTR (**Table S5**) and interact with the graphene sponge electrode surface via Van der Waals forces, in addition to π-π interactions. TCP is negatively charged at pH 7 and may undergo electrostatic interactions with the graphene sponge anode, in addition to π-π interactions with both electrodes. Thus, electrochemical degradation of CBZ, DEET and TCP also involved $HO^{\bullet}_{ads}$ generated at both graphene sponge anode and cathode. It should be noted here that while saturated alcohols such as

*t*-BuOH have negligible reactivity with $^1O_2$ [51,52], thus making *t*-BuOH a selective scavenger for HO• species, AA undergoes Schenck ene reaction with $^1O_2$ to form 1,2-diols [64]. Thus, addition of AA can scavenge both surface-bound HO• and $^1O_2$, although it is expected to have lower reactivity with $^1O_2$ compared to FFA due to the presence of two C=C bonds and more substituted allylic H-atoms in FFA [65,66].

In the presence of 5 μM FFA, the removal efficiencies of IPM and DTR decreased from 95% to 56% (IPM) and 89% to 64% (DTR) (**Figure 6B**), confirming the role of dissolved, meta-stable $^1O_2$ in the electrooxidation of these more polar contaminants. However, in the case of TCP, addition of FFA had no impact on its removal, as expected from the well-known poor reactivity of electron-deficient contaminants with $^1O_2$ [51–53]. Addition of low concentrations of FFA decreased the removal efficiency of CBZ from 71 to 48%, whereas the removal of DEET was decreased from 31 to 24%. In summary, $^1O_2$ is expected to contribute to a significant degree to the electro-oxidation of polar contaminants, e.g., IPM and DTR, which also have electron-rich secondary and tertiary amine groups, via a non-radical pathway. For less polar contaminants without electron-donating substituents at the aromatic moieties, lower contribution of $^1O_2$ can be expected. Electrochemically generated $^1O_2$ will not contribute to the oxidation of contaminants that are electron-deficient, such as TCP.

## Conclusions

Graphene sponge anodes functionalized with 2D materials offer higher material conductivity, improved electron transfer, enhanced generation of strong oxidants (HO•, $O_3$) and improved electrochemical degradation of persistent organic contaminants IPM, DTR, CBZ, TCP and DEET.

Depending on the nature of the incorporated 2D material, the material can be tailored to promote its surface interaction with specific contaminants. For instance, hBN-RGO anode exhibited higher reactivity with less polar contaminants, CBZ and DEET, due to the enhanced π-π interactions. In the case of borophene-RGO anode, enhanced formation of oxidant species, namely ozone and HO•, improved the degradation of highly polar contaminants IPM and DTR. Experiments with specific radical scavengers demonstrated the key role dissolved species $HO\bullet_{dis}$ and $^1O_2$ in electrochemical degradation of highly polar iodinated contrast media IPM and DTR. Less polar contaminants (CBZ, DEET, TCP) that interact more with the graphene sponge surface via van der Waals, π-π and electrostatic interactions, will be primarily degraded by the surface-bound $HO\bullet_{ads}$ and $^1O_2$ (CBZ, DEET). This study investigated for the first time anodic catalysis of graphene-based electrodes doped with 2D materials for the degradation of organic contaminants, demonstrating that functionalization of RGO coating with trace amounts of 2D materials is a plausible strategy for improving the electrochemical removal of persistent organic contaminants through their enhanced interaction with the graphene sponge surface and enhanced generation of oxidant species.

## Acknowledgments

The authors would like to acknowledge ERC Starting Grant project ELECTRON4WATER (Three-dimensional nanoelectrochemical systems based on low-cost reduced graphene oxide: the next generation of water treatment systems), project number 714177. ICRA researchers thank funding from CERCA program.

# Supplementary Material

## Functionalization of graphene sponge electrodes with two-dimensional materials for tailored electrocatalytic activity towards contaminants of emerging concern

Elisabeth Cuervo Lumbaque [a,b], Luis Baptista-Pires[a,b], Jelena Radjenovic[a,c]

[a]*Catalan Institute for Water Research (ICRA), Emili Grahit 101, 17003 Girona, Spain*

[b]*University of Girona, Girona, Spain*

[c]*Catalan Institution for Research and Advanced Studies (ICREA), Passeig Lluís Companys 23, 08010 Barcelona, Spain*

* *Corresponding author:*

*Jelena Radjenovic, Catalan Institute for Water Research (ICRA), Emili Grahit 101, 17003 Girona, Spain*

Phone: + 34 972 18 33 80; Fax: +34 972 18 32 48; E-mail: jradjenovic@icra.cat

**Text S1. Surface characterization of graphene-based sponges.**

A FEI Quanta FEG (pressure: 70Pa; HV: 20kV; and spot: 4) was used for the scanning electron microscopy (**SEM)** analysis of the morphology of the sponges and confirm the presence of 2D materials. X-ray photoelectron spectroscopy (**XPS)** measurements were carried out in ultra-high vacuum (base pressure 1-10 mbar) using a Phoibos 150 analyzer (SPECS GmbH, Germany) with a monochromatic aluminium Kalpha x-ray source (1486.74 eV). The energy resolution as measured by the FWHM of the Ag 3d5/2 peak for a sputtered silver foil was 0.58 eV. X-ray diffractometry **(XRD)** data was acquired with an X'pert multipurpose diffractometer at room temperature using a Cu Kα radiation (l = 1.540 Å). The detector used is an X'Celerator that is an ultrafast X-ray detector based on a real time multiple strip technology. The diffraction pattern was recorded between 20° and 70° using a step size of 0.03° and a time per step of 1,000 s. **Zeta potential** of RGO, borophene-RGO and hBN-RGO was measured using Zetasizer Nano ZS (Malvern Panalytical Ltd) operating with a 633 nm laser and using an aqueous solution at pH 7. Multi-channel potentiostat/galvanostat VMP-300 (BioLogic, U.S.A.) was used for the electrochemical impedance spectroscopy **(EIS)** and cyclic voltammetry **(CV)** analysis (range 0-0.8V for capacitance analysis and 0-2.5V for system monitoring).

**Text S2. Analytical methods**

Target organic contaminants were analyzed with a 5500 QTRAP hybrid triple quadrupole linear ion trap mass spectrometer (QLIT-MS) with a turbo Ion Spray source (Applied Biosystems), coupled to a Waters Acquity Ultra-PerformanceTM liquid chromatograph (UPLC). Iopromide (IPM), diatrizoate (DTR), carbamazepine (CBZ) and N,N-Diethyl-m-toluamide (DEET) were analyzed in electrospray (ESI) positive mode using an Acquity UPLC HSS T3 column (2.1×50 mm, 1.8 μm, Waters) run at 30°C. The eluents employed were acetonitrile with 0.1% formic acid (eluent A), and Milli-Q water with 0.1% formic acid (eluent B) at a flow rate of 0.5 mL $min^{-1}$. The gradient was started at 2% of eluent A that was increased to 20% A by 3 min, further increased to 50% A by 6 min and further increased to 95% A by 7 min. It was kept constant for 2.5 min, before returning to the initial condition of 2% A by 9.5 min. The total run time was 11 min. Triclopyr (TCP) was analyzed in the ESI negative mode using an Acquity UPLC BEH C18 column (2.1×50mm, 1.7μm) from Waters run at 30°C. The eluents for ESI negative mode were mixture of acetonitrile and methanol (1:1, v/v) (eluent A) and 1 mM ammonium acetate (eluent B) at a flow rate of 0.6 mL $min^{-1}$. The gradient was started at 5% A, further increased to 100% A by 7 min and then kept constant for 2 min, before returning to the initial conditions of 5 % A by 9 min. The total run time in the ESI negative mode was 10 min.

The target organic contaminants were analyzed in a multiple reaction monitoring (MRM). The source-dependent parameters were as follows: for the positive mode; curtain gas (CUR), 30 V; nitrogen collision gas (CAD), medium; source temperature (TEM), 650°C; ion source gases GS1, 60 V and GS2, 50 V; ion spray voltage, 5500V, and entrance potential (EP), 10V. For the negative mode; curtain gas (CUR), 30V; nitrogen collision gas (CAD), medium; source temperature (TEM), 650°C; ion source gases GS1, 60 V and GS2, 70 V; ion spray voltage, -3500 V, and entrance potential (EP), -10V. The optimized compound-dependent MS parameters for each compound are summarized in **Table S1**.

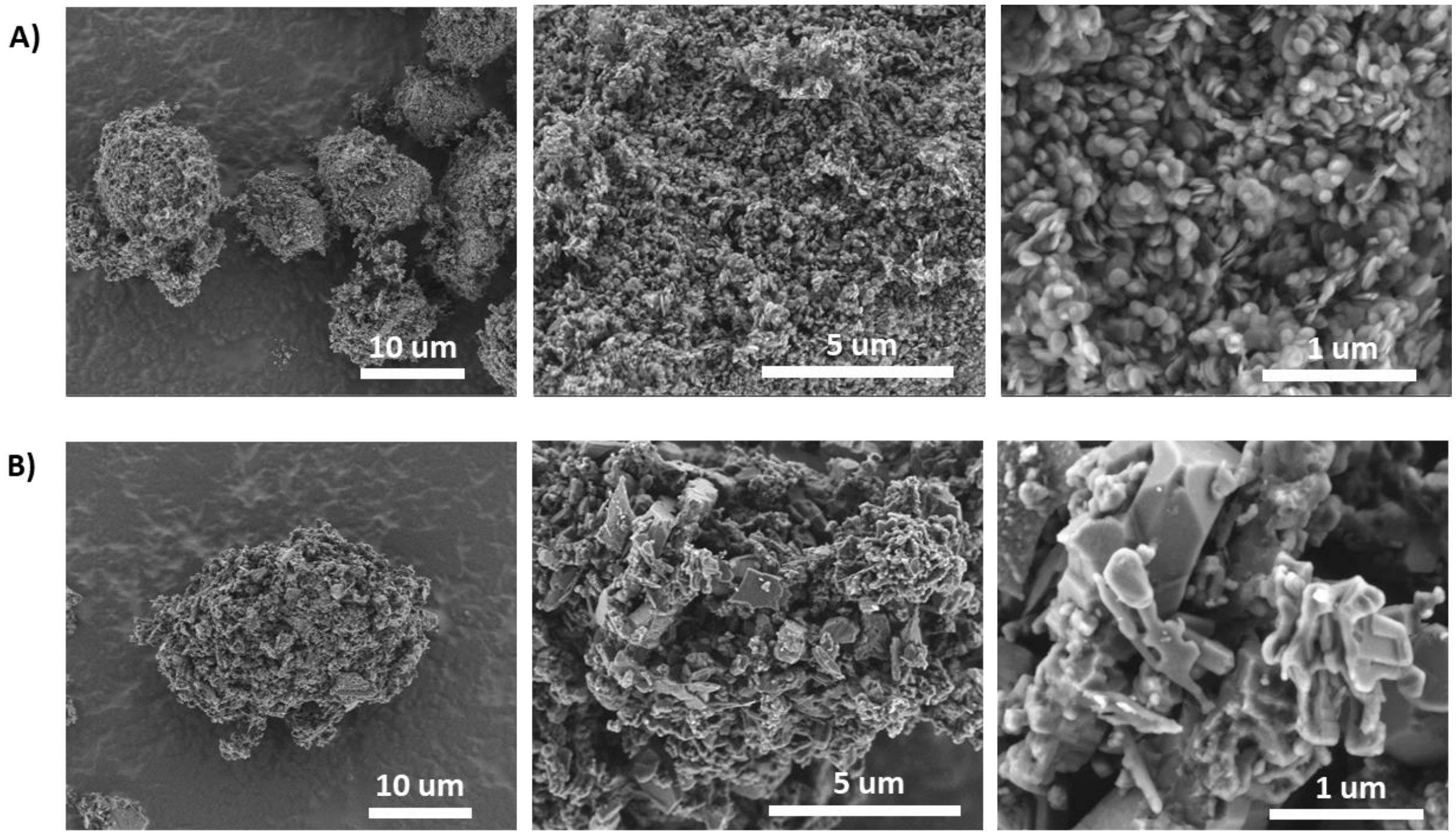

**Figure S1.** Scanning electron microscopy (SEM) analysis of **A)** bulk boron nitride and **B)** bulk boron.

A)

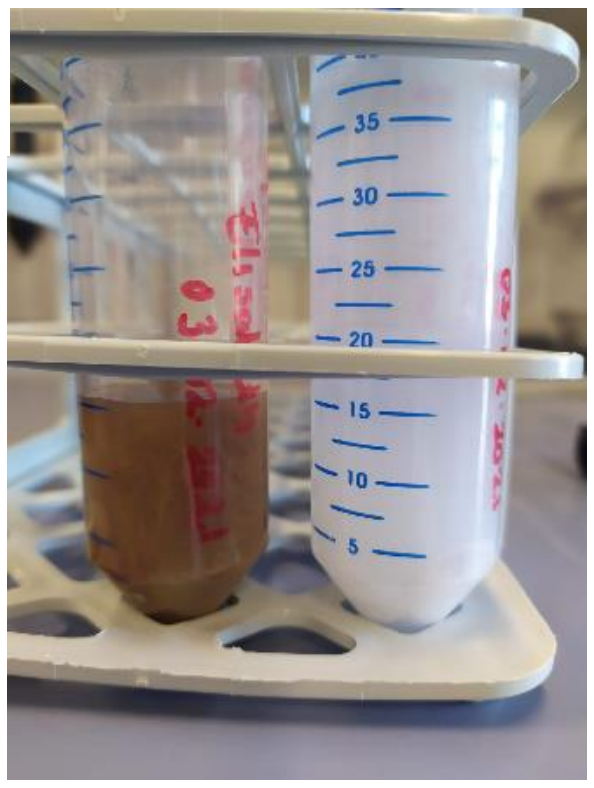

B)

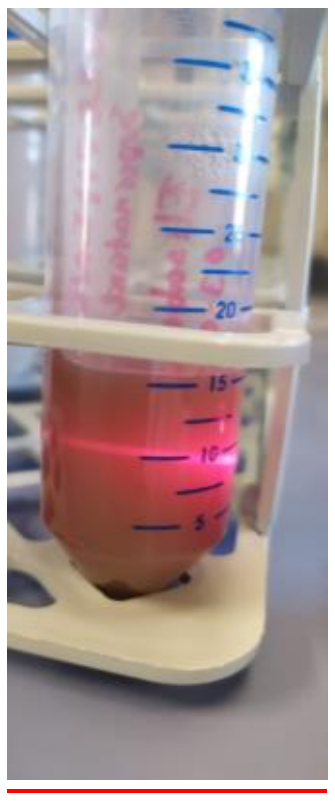

C)

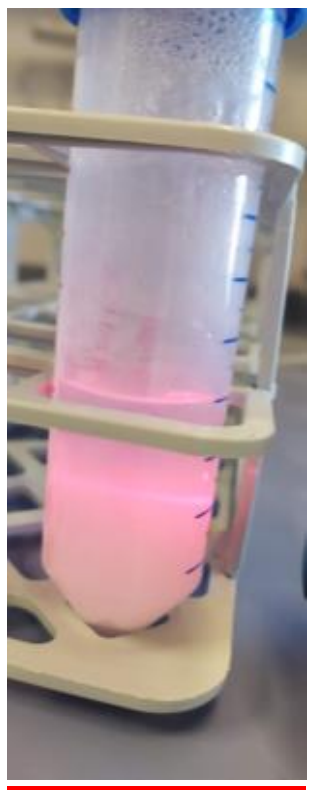

**Figure S2.** Tyndal effect of: **A)** boron (brown) and boron nitride (white) aqueous solution, **B)** borophene dispersion, and **C)** hexagonal boron nitride (hBN) dispersion.

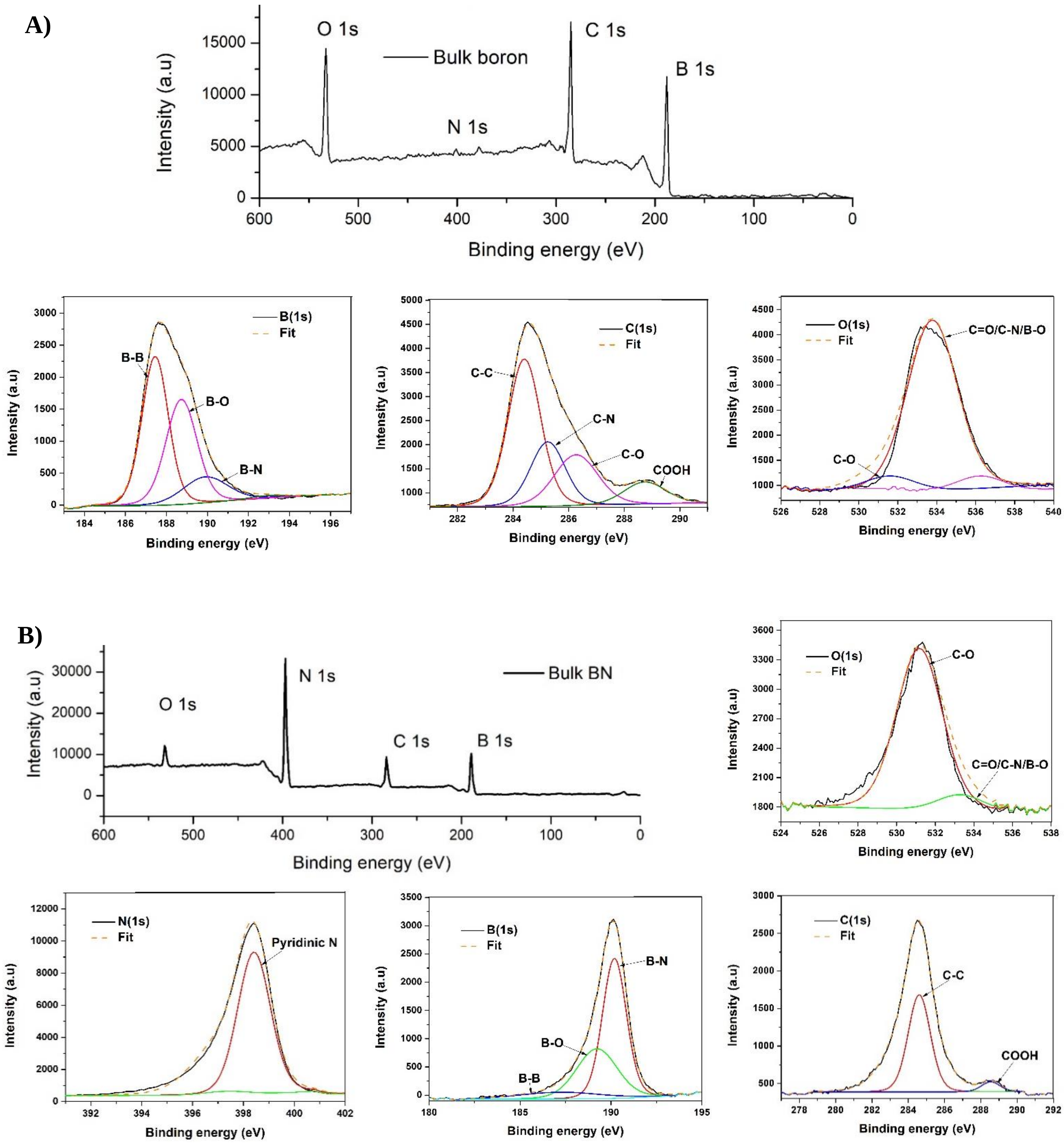

**Figure S3.** XPS of **A)** boron and **B)** boron nitride.

A)

B)

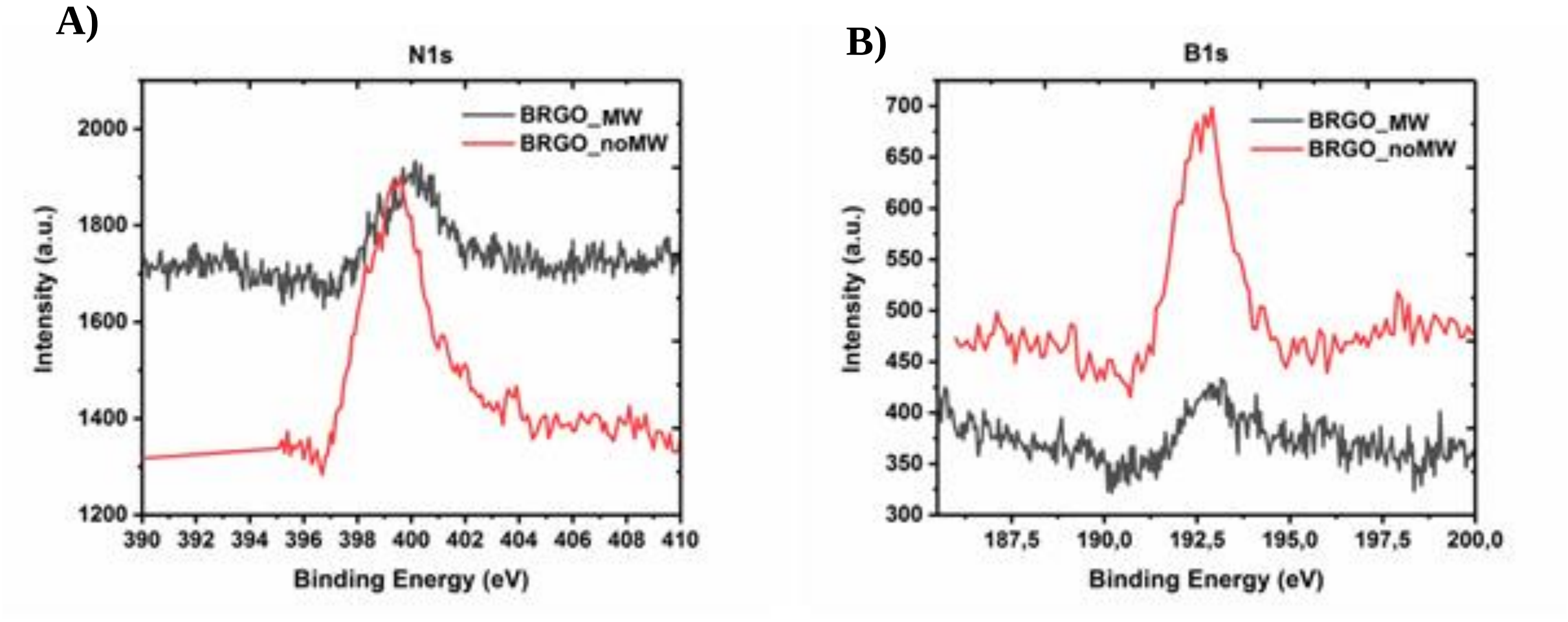

**Figure S4.** XPS of **A)** N1s and **B)** B1s spectra of B-RGO coating on a mineral wool substrate (MW) and without mineral wool (no MW).

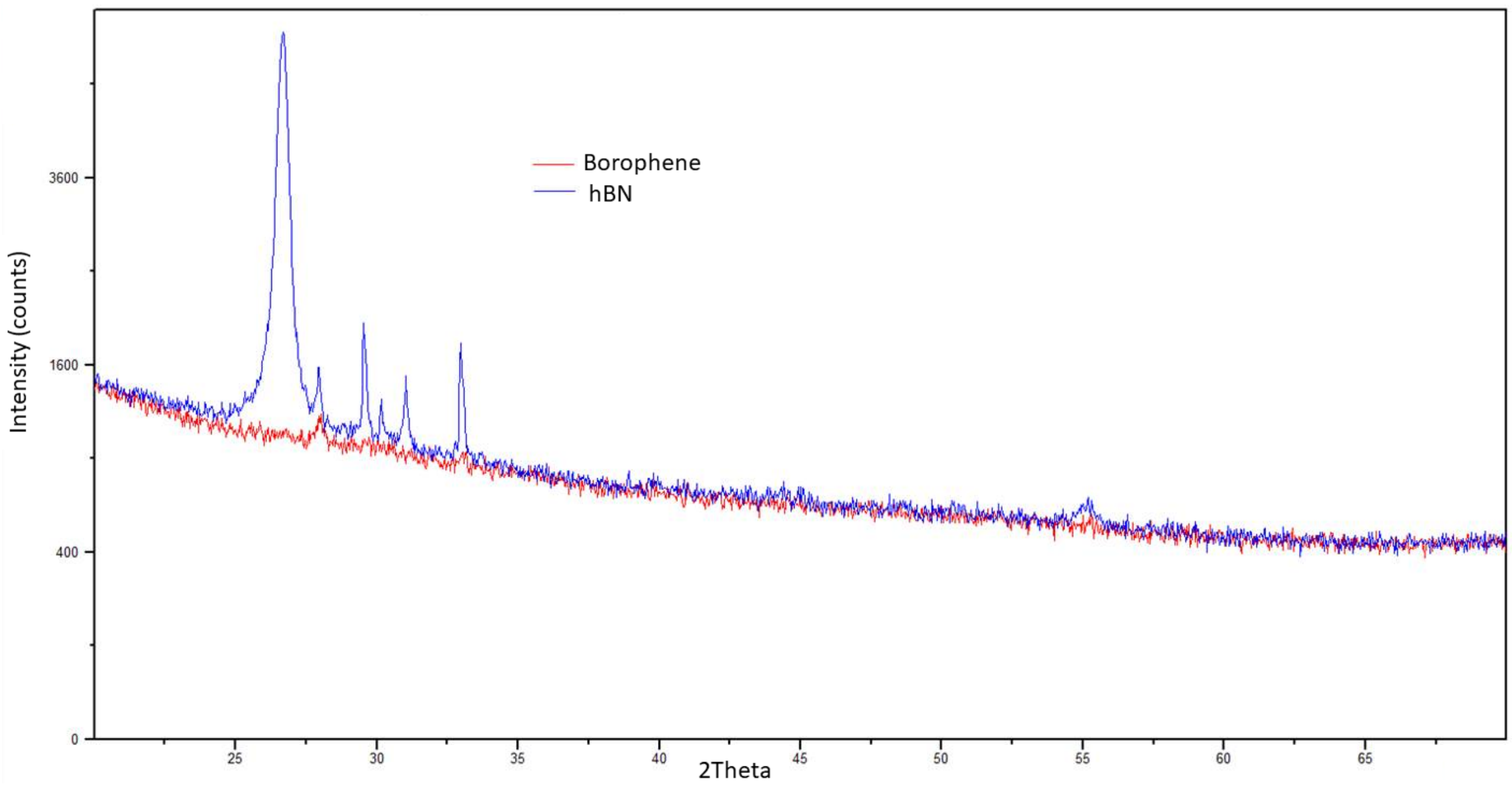

**Figure S5.** XRD of supernatants: **A)** borophene and **B)** hBN.

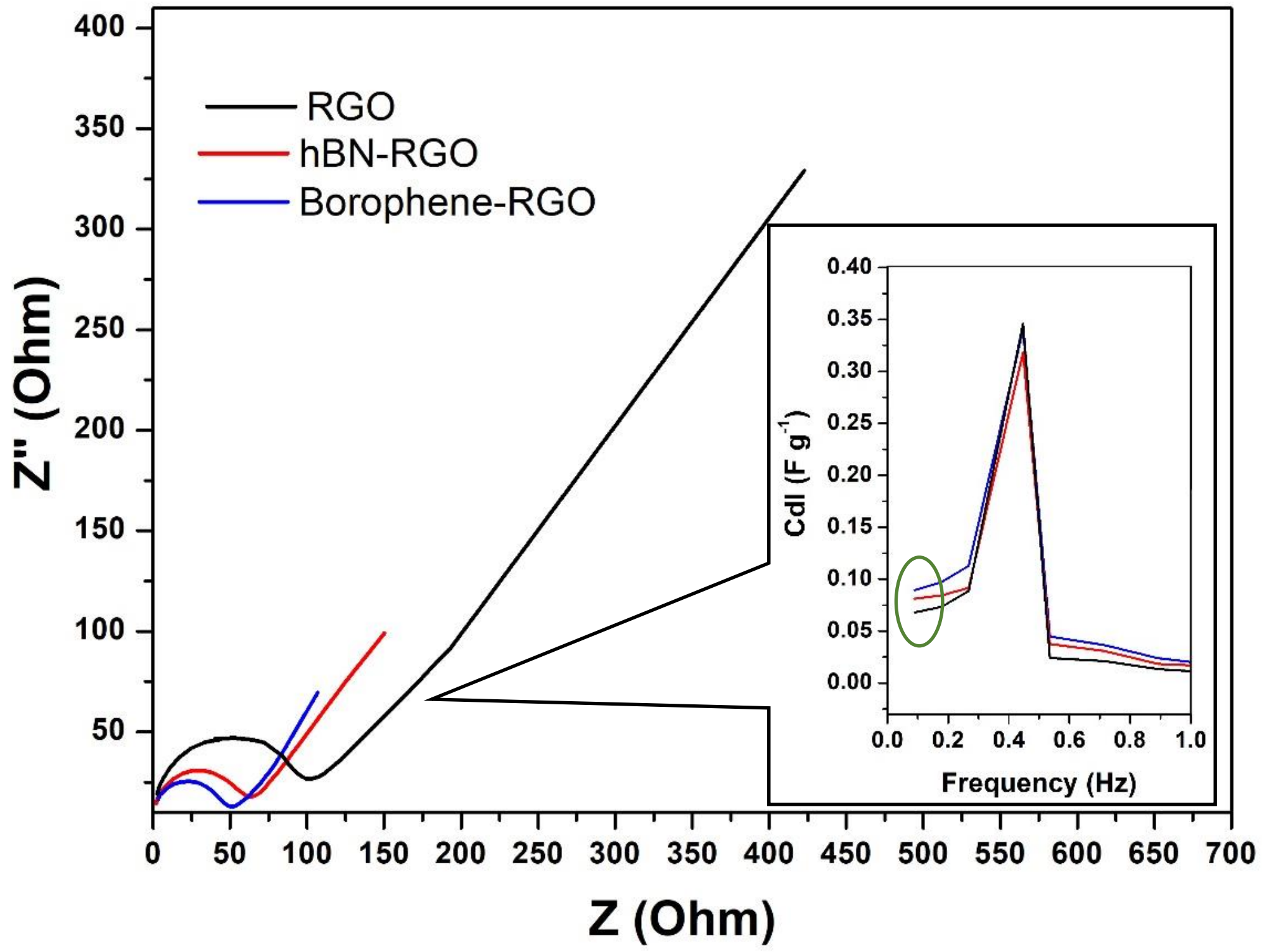

**Figure S6. A)** Nyquist plots of anodes sponges, three experiments in 1 M phosphate buffer, conductivity 46 mS $cm^{-1}$ and **B)** imaginary part of the capacitance as a function of the frequency.

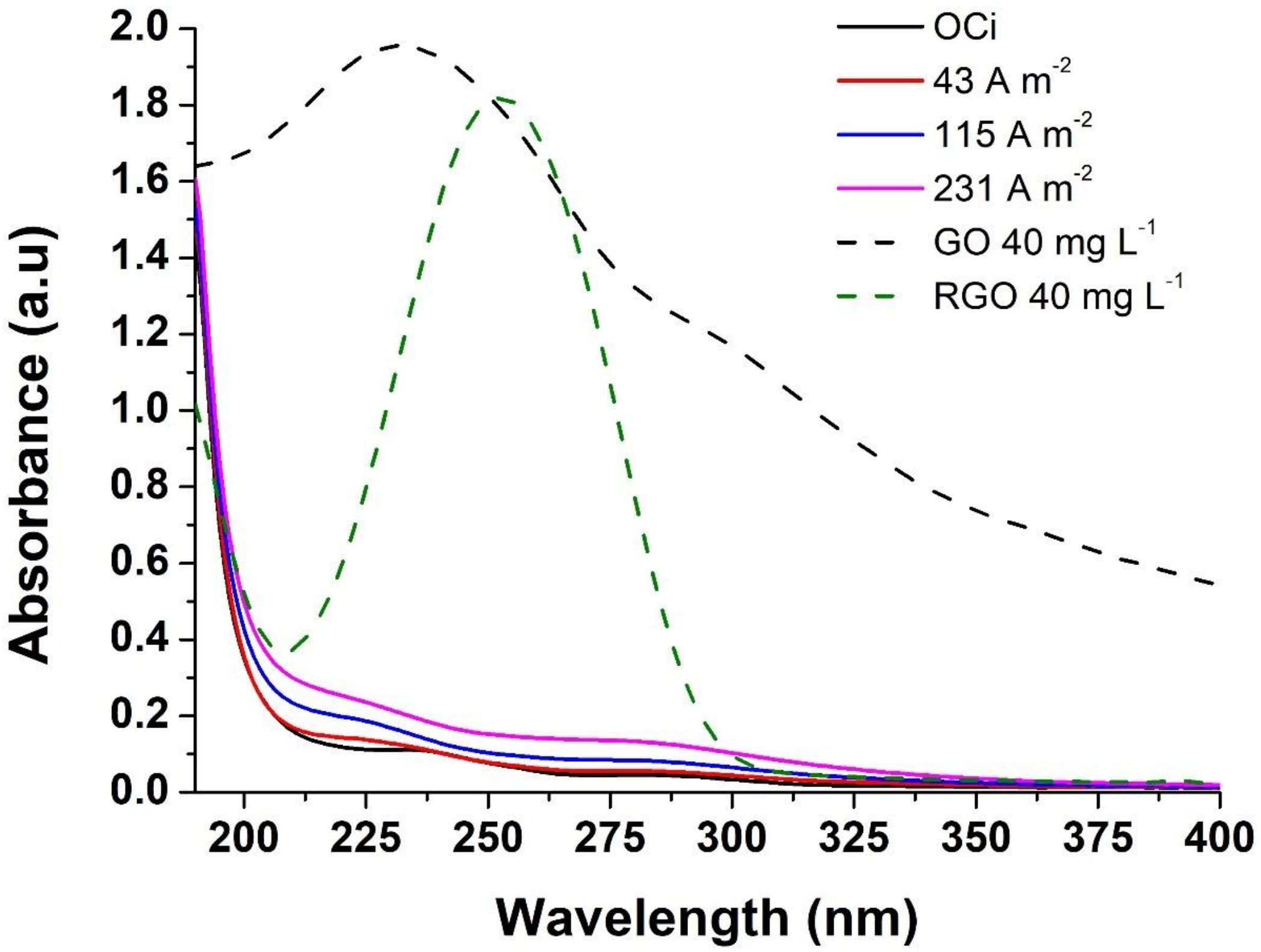

**Figure S7.** Absorption spectra of GO solution at 40 mg $L^{-1}$ (black-dotted line), RGO dispersion (green-dotted line) and effluent samples taken after the open circuit (OCi) and chronopotentiometric experiments at 43, 115 and 231 A $m^{-2}$. GO displays an absorption peak at 230 nm in accordance with the literature [1,2]. To illustrate a shift in absorbance of GO to the higher wavelengths, RGO dispersion was prepared using milder reduction procedure to facilitate the re-suspension of RGO and absorbance measurement. Given that this reduction was incomplete, the absorbance peak was located at 254 nm, instead of 268-270 nm reported in literature [1,2].

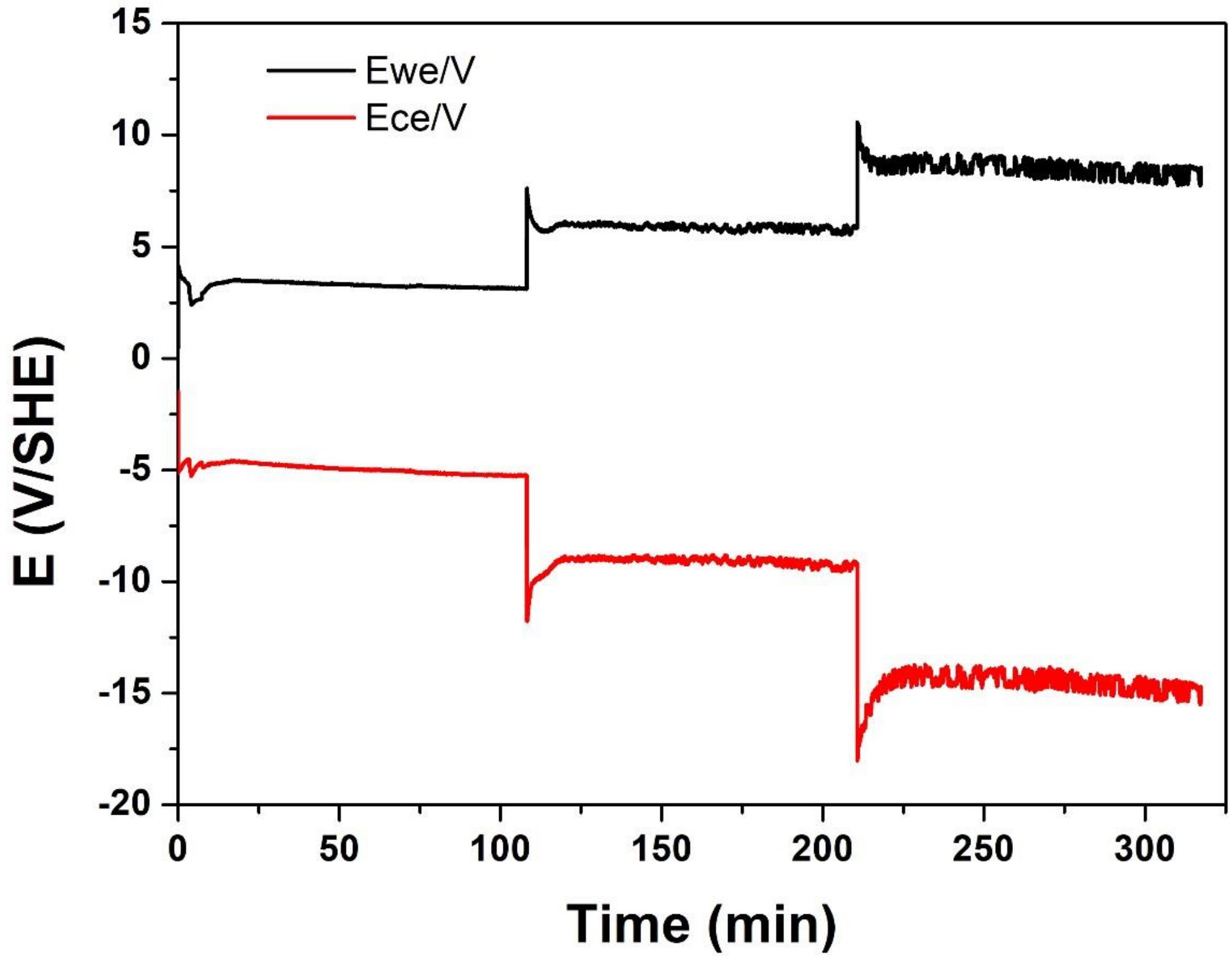

**Figure S8.** Recorded potentials of anode (working electrode, Ewe) and cathode (counter electrode, Ece) at different current densities (43, 115 and 231 A $m^{-2}$) for hBN-RGO anode – N-RGO cathode system.

**Table S1.** The optimized compound-dependent MS parameters: declustering potential (DP), collision energy (CE) and cell exit potential (CXP) for each compound and each transition of the negative and positive mode.

| Organic compound | Q1 Mass (Da) | Q3 Mass (Da) | DP | CE | CXP |
|---|---|---|---|---|---|
| **Diatrizoate** | 614.9 | 361 | 80 | 30 | 10 |
| | 614.9 | 233.1 | 85 | 33 | 10 |
| **Carbamazepine** | 237.01 | 194.1 | 156 | 47 | 10 |
| | 231.01 | 193 | 156 | 47 | 10 |
| **Iopromide** | 791.72 | 572.9 | 156 | 35 | 20 |
| | 791.72 | 300.0 | 156 | 83 | 10 |
| **Triclopyr** | 256.18 | 197.7 | -75 | -16 | -5 |
| | 254.01 | 196 | -55 | -30 | -10 |
| **DEET** | 192.3 | 119.3 | 176 | 23 | 14 |
| | 192.3 | 91.6 | 176 | 39 | 12 |

**Table S2.** Limits of detection (LOD) and quantification (LOQ) of model contaminants.

| | LOD (μM) | LOQ (μM) |
|---|---|---|
| **Diatrizoate (DTR)** | 0.004 | 0.01 |
| **Carbamazepine (CBZ)** | 0.03 | 0.09 |
| **Iopromide (IPM)** | 0.006 | 0.02 |
| **Triclopyr (TCP)** | 0.002 | 0.008 |
| **N,N-Diethyl-meta-toluamide (DEET)** | 0.04 | 0.1 |

**Table S3.** XPS atomic content of the GO precursor solution, graphene sponge (RGO), borophene and hBN precursor solutions and supernatants

| | **GO** | **RGO** | **Boron** | **Borophene supernatant** | **Boron nitride** | **hBN supernatant** |
|---|---|---|---|---|---|---|
| **C (%)** | 62.6 | 70.3 | 31.9 | 21.9 | 16.3 | 6.4 |
| **O (%)** | 36.9 | 28 | 12.4 | 32.1 | 5.3 | 4.4 |
| **N (%)** | 0.9 | 1.7 | 1.2 | 1.5 | 38.6 | 44.3 |
| **B (%)** | 0 | 0 | 54.3 | 44.5 | 39.9 | 44.9 |

**Table S4.** XPS atomic content analysis of the samples: percentage of functional groups of C1s, O1s, N1s and B1s XPS spectra of the synthetized materials.

| | C1s (%) | | | | O1s (%) | | | N1s (%) | | | | B1s (%) | | | |
|---|---|---|---|---|---|---|---|---|---|---|---|---|---|---|---|
| | *C-C 284.5eV* | *C-N 285.6eV* | *C-O 286.9eV* | *COOH 288.5eV* | *C-O 531eV* | *C=O/ C-N/B-O 533eV* | *536.3eV adsorbed water and/or oxygen* | *Pyridinic N 398.4eV* | *Pyrrolic N 399.8eV* | *Graphitic-N 401.6eV* | *Azide 402.7eV* | *B-O 188.9 eV* | *B-B 187.6 eV* | *B-N 190.9 eV* | *B2O3 192.9 eV* |
| **GO** | 26.3 | 0 | 31.9 | 4.4 | 35.4 | 23.2 | 0 | 0 | 0 | 0.1 | 0 | | | | |
| **RGO** | 30.3 | 30.9 | 9.8 | 0 | 7.,8 | 16.5 | 3.6 | 0.4 | 0.7 | 0.,6 | 0 | | | | |
| **BN** | 12.3 | 0.8 | 1.9 | 1.3 | 4.2 | 0.8 | 0.19 | 34.7 | 2.9 | 0.8 | 0 | 6.1 | 3.2 | 29.2 | 1.4 |
| **hBN supernatant** | 6.1 | 0.16 | 0.16 | 0 | 2.2 | 2.2 | 0 | 42.1 | 2.2 | 0 | 0 | 39.8 | 1.8 | 1.8 | 1.5 |
| **Boron** | 20.1 | 2.7 | 5.8 | 3.3 | 0.7 | 11.1 | 0.5 | 0.2 | 0.1 | 0.9 | 0 | 5.0 | 45.9 | 2.6 | 0.8 |
| **Borophene supernatant** | 12.2 | 3.5 | 4.2 | 1.9 | 7.5 | 24.3 | 0.4 | 0.05 | 0.4 | 0.8 | 0.2 | 11.4 | 30.7 | 2.4 | 0.1 |

**Table S5.** Chemical structures with represented charge at neutral pH, and physico-chemical properties of the target contaminants: molecular weight (MW), pKa, octanol-water distribution coefficient calculated based on chemical structure at pH 7.4 (ACD/logD), and polar surface area. The information was obtained from https://chemicalize.com/

| Organic compound MW (g mol$^{-1}$) | Chemical structure | pka | logD | Polar surface area (Å$^2$) |
|---|---|---|---|---|
| Iopromide (791.11) | 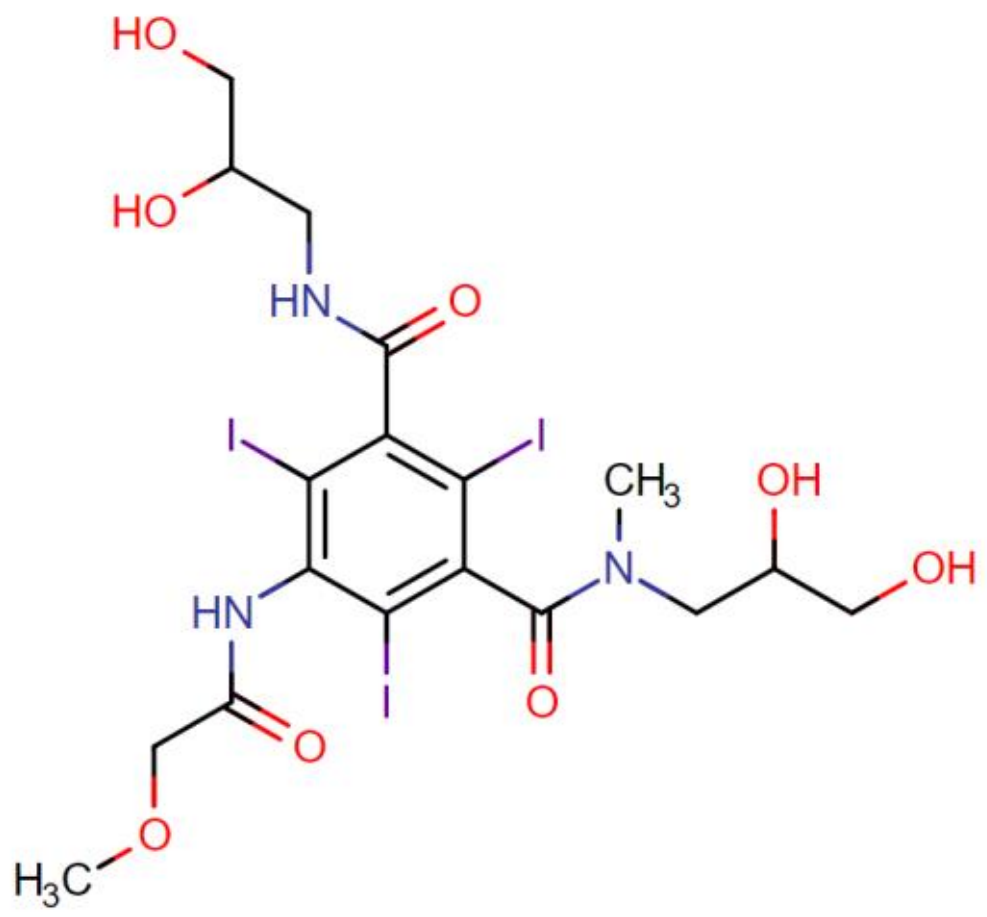 | 11.1 | -2.1 | 169 |
| Diatrizoate (613.91) | 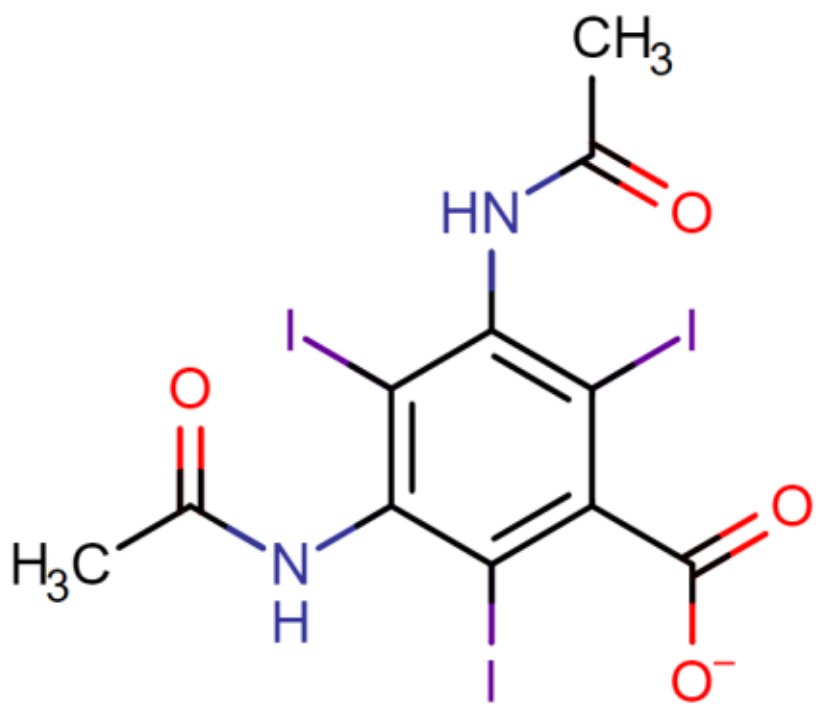 | 2.2 | -2.1 | 96 |
| Carbamazepine (236.27) | 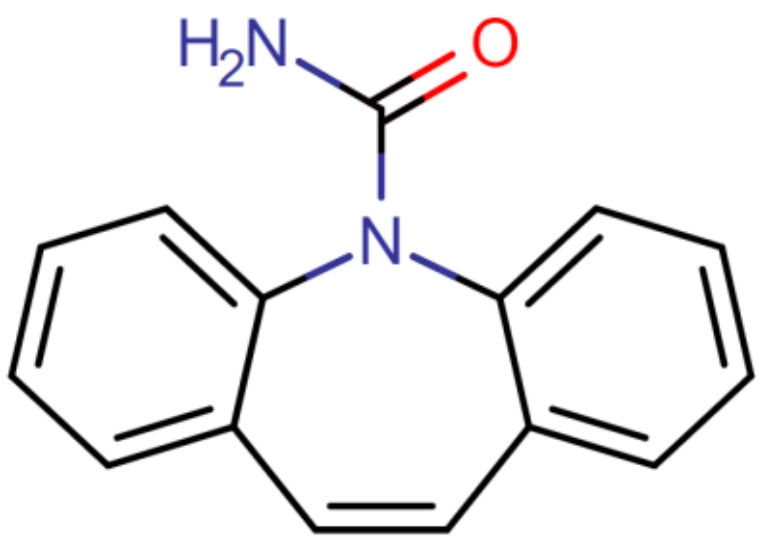 | 15.96 | 2.3 | 46 |
| Triclopyr (256.46) | 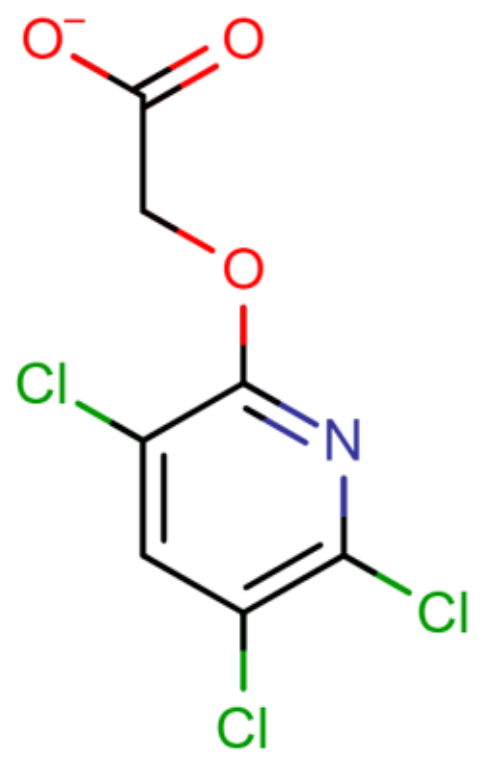 | 2.28 | -1.0 | 59 |
| DEET (191.27) | 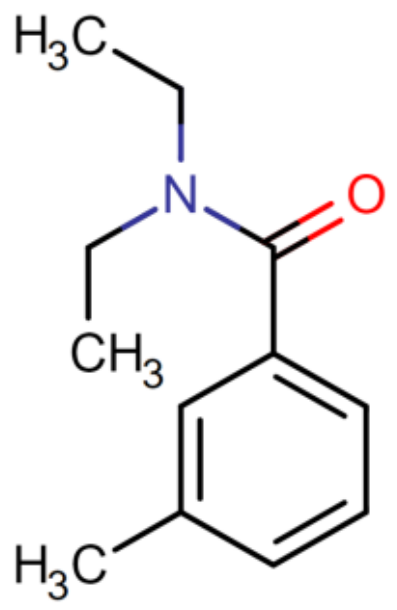 | -0.98 | 2.2 | 20 |

**Table S6.** Recorded potentials (V vs Standard Hydrogen Electrode, /SHE) for anode ($E_{AN}$) and cathode ($E_{CAT}$), and total cell potentials ($E_{TOT}$, V) at different current densities (J, A $m^{-2}$) for the three employed anode (A) – cathode (C) electrochemical systems.

| System | J, A $m^{-2}$ | $E_{AN}$, V/SHE | $E_{CAT}$, V/SHE | $E_{TOT}$, V |
|---|---|---|---|---|
| **borophene-RGO (A) – N-RGO (C)** | **43** | 3±0.1 | -6.5±0.2 | 9.5±0.2 |
| | **115** | 4±0.5 | -11±0.5 | 15±0.5 |
| | **231** | 6±2 | -15.5±2 | 21±2 |
| **hBN-RGO (A) – N-RGO (C)** | **43** | 4±0.1 | -5±0.9 | 9±1.0 |
| | **115** | 5±1.4 | -9.3±0.9 | 14.3±1.4 |
| | **231** | 7±2.4 | -14±0.1 | 21.2±2.4 |
| **RGO (A) – N-RGO (C)** | **43** | 4.1±0.7 | -4.7±0.4 | 8.8±0.4 |
| | **115** | 6.2±0.4 | -8±0.5 | 14.2±0.4 |
| | **231** | 8.2±0.1 | -13.8±0.7 | 22±0.8 |

**Table S7.** Ohmic-drop corrected potentials for RGO, borophene-RGO and hBN-RO anode vs Standard Hydrogen Electrode (SHE) at different currents for the three electrochemical systems.

| | **Recorded anodic potential (V/SHE)** | | | **Ohmic-drop corrected anodic potential (V/SHE)** | | |
|---|---|---|---|---|---|---|
| | **43 $Am^{-2}$** | **115 $Am^{-2}$** | **231 $Am^{-2}$** | **43 $Am^{-2}$** | **115 $Am^{-2}$** | **231 $Am^{-2}$** |
| **RGO** | **4.1** | **6.2** | **8.2** | **3.6** | **4.8** | **5.3** |
| **Borophene-RGO** | **3** | **4** | **6** | **2.5** | **2.6** | **3.1** |
| **hBN-RGO** | **4** | **5** | **7** | **3.5** | **3.6** | **4.3** |

**Table S8.** Reported bimolecular rate constants of target contaminants with ozone and hydroxyl radicals at neutral pH.

| **Compound** | **$K_{O3}$ ($M^{-1}$ $s^{-1}$)** | **$k_{HO•}$ ($M^{-1}$ $s^{-1}$)** |
|---|---|---|
| Carbamazepine | $9.1 \times 10^5$ [3] | $9.1 \times 10^9$ [4] |
| Iopromide | 93.07 [5] | $3.34 \times 10^9$ [6] |
| Diatrizoate | 48.65 [5] | $9.58 \times 10^8$ [6] |
| Triclopyr | 105 [7] | 1.7 x $10^9$ [7] |
| DEET | 0.126 [8] | $4.95\times 10^9$ [8] |